\documentclass[%
 reprint,
superscriptaddress,
 amsmath,amssymb,
 aps,
nofootinbib]{revtex4-2}

\usepackage{color}
\usepackage{hyperref}

\usepackage{graphicx}
\usepackage{dcolumn}
\usepackage{bm}

\begin{document}

\preprint{APS/123-QED}

\title{ Cosmological perturbation theory using generalized Einstein de Sitter cosmologies}


\author{Michael Joyce}\email{joyce@lpnhe.in2p3.fr}
\affiliation{Laboratoire de Physique Nucléaire et de Hautes Energies, UPMC IN2P3 CNRS UMR 7585, Sorbonne Université, 4, place Jussieu, 75252 Paris Cedex 05, France}
\author{Azrul Pohan}\email{azrul.pohan@lpnhe.in2p3.fr}
\affiliation{Laboratoire de Physique Nucléaire et de Hautes Energies, UPMC IN2P3 CNRS UMR 7585, Sorbonne Université, 4, place Jussieu, 75252 Paris Cedex 05, France}
\affiliation{Laboratory of Theoretical Physics, Institut Teknologi Sumatera, Lampung 35365, Indonesia}




\begin{abstract}
The separable analytical solution in standard perturbation theory for an Einstein de Sitter (EdS) universe can be generalized to the wider class of such cosmologies (``generalized EdS'', or gEdS)  in which a fraction of the pressure-less fluid does not cluster. We derive the corresponding kernels in both Eulerian perturbation theory (EPT) and Lagrangian perturbation theory, generalizing the canonical EdS expressions to a one-parameter family where the parameter can be taken to be the exponent $\alpha$ of the growing mode linear amplification $D(a) \propto a^{\alpha}$.  For the power spectrum (PS) at one loop in EPT,  the contribution additional to standard EdS is given, for each of the `13' and `22' terms, as a function of two infra-red safe integrals. In the second part of the paper we show that the calculation of cosmology-dependent corrections in perturbation theory in standard (e.g. LCDM-like) models can be simplified, and their magnitude and parameter dependence better understood, by relating them to our analytic results for gEdS models. At second order the time dependent kernels are equivalent to the analytic kernels of the gEdS model with $\alpha$ replaced by a single redshift dependent effective growth rate $\alpha_2(z)$.  At third order 
the time evolution can be conveniently parametrized in terms of two additional such  effective growth rates. For the PS calculated at one loop order, the correction to the PS relative to the EdS limit can be expressed in terms of just 
$\alpha_2(z)$, one additional effective growth rate function and  the four infra-red safe integrals of the gEdS limit. This is much simplified compared to expressions in the literature that use six or eight red-shift dependent functions and are not explicitly infra-red safe. Using the analytic gEdS expression for 
the PS with $\alpha=\alpha_2(z)$ gives a good  approximation (to $\sim 25 \%$) for the exact result.
\end{abstract}
\maketitle

\section{Introduction}
\label{sec:intro}
Understanding the origin of the large-scale structures in the Universe first revealed several decades ago by 
early red-shift surveys remains a central and open problem in cosmology. This is driven by the ever-increasing richness and accuracy of observational data which will grow in the coming years with programs such as Euclid, DESI and LSST \cite{laureijs2011euclid,dey2019overview,ivezic2019lsst}.
In standard cosmological models the clustering observed at cosmological scales today is the result of evolution under the gravity of primordial fluctuations,
that is highly constrained by observations of the fluctuations in the cosmic microwave background, with 
potential fluctuations $\sim 10^{-5}$ at the time of decoupling. Calculation of predictions derived from the non-linear equations governing the evolution of matter fluctuations rests in practice essentially on numerical solutions using the $N$-body method (for a recent review, see e.g. \cite{Angulo2021}). At sufficiently early times and/or large scales, when fluctuations to uniformity are small, analytical perturbative approaches may be used and provide an invaluable guideline and control on numerical simulations. Perturbation theory beyond the leading (linear) order was originally treated in early works (e.g. \cite{Juszkiewicz1981, Vishniac1983, Goroff1986, Suto1991, Bertschinger1994,bernardeau1993nonlinear,bernardeau1993skewness,catelan1995eulerian,makino1992,Bernardeau2002}) and has been 
developed since in an extensive literature (for a review, see  \cite{Bernardeau2002}). In recent years perturbation theory has remained highly topical, with development of various different aspects and cosmological applications (see e.g. \cite{crocce2006renormalize,CarlsonPRD2009,taruya2016constructing,ModiCMB2017,chen2020consistent,garny2021loop,ChenLyman2021,ChenLPT2022}), and, more recently, of the so-called Effective Field Theory (EFT) of cosmological structure formation (see e.g. \cite{crocce2006renormalize, mcdonald2007dark, carrasco2012effective, pajer2013renormalization, steele2021precise}).

In this article we explore and derive results for the perturbation theory of cosmological structure formation in a particular set of idealized cosmological models. We refer to these as  ``generalized Einstein de Sitter'' (gEdS) models as they are models in which the background cosmological evolution is that of an EdS cosmology. However, only a fraction of the total energy density driving the expansion of the background corresponds to clustering matter, the rest remaining uniform. This is simply equivalent to a smooth matter-like component (e.g. arising from a homogeneous mode of a scalar field, or from a massive neutrino-like contribution when perturbations are neglected). We derive analytical solutions for the key quantities (kernels) in both standard (Eulerian) perturbation theory (SPT) and also in Lagrangian perturbation theory (LPT), and use them to derive the expression for the power spectrum at the leading non-trivial order
(one loop). The motivation for studying these models is twofold. On the one hand, as we will show in the second part of this paper, these models turn out to be very useful to better understand, and simplify the calculation of, the small cosmology-dependent corrections to results obtained using the usual ``separable EdS''  approximation made in applying perturbation theory to the standard (i.e. LCDM-like) cosmologies \cite{ takahashi2008third, fasiello2016nonlinear, Aviles2018Nonlinear,baldauf2021two, garny2021loop, alkhanishvili2022reach,fasiello2022perturbation,Bernardeau2002}. 
In particular we will show how expressions for the non-EdS corrections to the one-loop PS in 
perturbation theory can be given (exactly) in terms of just two redshift dependent 
``effective growth rates'', and, to a good approximation, in terms of a single 
such function. 
On the other hand, our motivation  comes from the intrinsic interest of new analytical results in perturbation theory in order to test its validity, and in particular to probe its accuracy in predicting the dependence on expansion history 
beyond the usually used approximation. The models we focus on are particularly suitable for robust numerical tests as in the case of initial Gaussian fluctuations specified by a simple power-law power spectrum these gEdS models are scale-free and characterized (at least for some range of spectra) by ``self-similarity'' in their evolution 
\cite{dbenhaiem_PhD, benhaiem2014self}. In principle this property allows one to obtain  very accurate numerical results with which to test perturbation theory (in particular by applying the methods recently discussed for scale-free EdS models in  \cite{joyce2021quantifying,maleubre2022accuracy}). An analysis of these ``generalized scale-free'' models and numerical tests will be reported in a separate forthcoming  article.

The article is organized as follows. The next section presents first the family of gEdS models, and then the calculation of  perturbation theory kernels in them for both Eulerian and Lagrangian formulations, giving the density and velocity kernels up to the third order. In section 3 we study the convergence properties of the one-loop PS in gEdS cosmologies, considering how the well-known results for standard EdS are modified. Section 4 considers how the functions characterising the exact evolution of the PS in perturbation theory in a class of standard (LCDM-like) cosmological models can be approximated by interpolation of gEdS models.  
In the final section we summarize and comment on further possible exploitation of the results we have derived, notably in relation to understanding better the ultra-violet divergences in standard perturbation theory. Some details of our calculations are relegated to appendices, as well as additional detailed comparison with the previous  results of \cite{ takahashi2008third}, and also to those of \cite{bernardeau1993skewness, garny2021loop,fasiello2022perturbation}. 
For readers wishing to make use of 
our simplified expression for the cosmological corrections to the EdS approximation for the one-loop PS in standard cosmologies, given
in terms of just two redshift-dependent functions,
without going through 
the full derivation given in the body of the article, we give the necessary 
equations also in a short appendix. 

\section{Perturbation Theory kernels in generalized EdS models}

\label{Perturbation Theory kernels in generalized EdS models}

\subsection{Generalized EdS models}
\label{Generalized EdS models}
We consider perturbed Friedman-Lemaitre-Robertson-Walker (FLRW) models with a 
pressureless perturbed matter component and a smooth component (i.e. without perturbations). 
The Friedmann equations 
can be written as    
\begin{equation}\label{gedsFried1}
\begin{split}
 \mathcal{H}^2&=\frac{8\pi G}{3}a^{2}\big[\rho_{m}+\rho_{s}\big],\\
  \frac{\partial{\mathcal{H}}}{\partial{\tau}}&=-\frac{4\pi G}{3}a^{2}\big[\rho_{m}+(1+3w_s)\rho_{s}\big],
\end{split}
\end{equation}
where $\mathcal{H}=d(\log a)/d\tau$ and $\tau$ is the conformal time related to cosmic time $t$ by $\tau=\int dt/a$, $a$ is the scale factor, $\rho_{m}$ is the mean density of the perturbed matter component, $\rho_{s}$ is the smooth ``dark energy'' component with equation of state $p_s=w_s \rho_s$ (where $w_s$ can be a function of $a$). The case $w_s=-1$ corresponds to the Lambda Cold Dark Matter (LCDM) cosmology. 
As canonically, we define the fraction of matter and dark energy as 
\begin{equation}
    \Omega_{m}=\frac{\rho_{m}}{\rho_{tot}},
    \quad 
    \Omega_{s}=\frac{\rho_{s}}{\rho_{tot}},
\end{equation}
where $\rho_{tot}=\rho_{m}+\rho_{s}$ is the total energy density (and $\Omega_{m}+\Omega_{s}=1$). Once $w_s$ is specified, the model is fully characterized by specifying the value at some time of one of these
two parameters. Our analysis of standard cosmologies below applies to 
this class of models, with the sole further assumption that the total energy density asymptotically scales as matter at early times. 

The family of cosmologies we will focus on first, those we will refer to as {\it generalized Einstein de Sitter cosmologies} (hereafter gEdS), simply correspond to the case $w_s=0$.  In this case 
\cite{benhaiem2014self, dbenhaiem_PhD} the smooth component behaves exactly like a matter component scaling as $\rho_{s} \propto 1/a^3$, so that both $\Omega_{m}$ and $\Omega_{s}$ are constant. They may thus be parameterized by this constant value of
$\Omega_{m}$ (or $\Omega_{s}$). In order to avoid confusion below with the general LCDM-type model where $\Omega_{m}$ is a function of time, we will use, as
in \cite{benhaiem2014self}, the parameter
\begin{equation}\label{kappa}
   \kappa^{2}=\frac{1}{\Omega_{m}} \,. 
\end{equation}
We then have that 
\begin{equation}
\begin{split}
 \mathcal{H}^2&=\frac{8\pi G}{3}a^{2}\kappa^2 \rho_{m},\\
  \frac{\partial{\mathcal{H}}}{\partial{\tau}}&=-\frac{4\pi G}{3}a^{2}\kappa^2 \rho_{m},
\end{split}
\end{equation}
and the scale factor $a(t)$ evolves as that of an EdS model, with
\begin{equation}\label{scale_factor}
\begin{split}
a=\Big(\frac{\kappa t}{t_{0}}\Big)^{2/3} \quad \text{where} \quad t_{0}=\frac{1}{\sqrt{6\pi G\rho_{m,0}}}\,,
\end{split}
\end{equation}
where $\rho_{m,0}$ is the matter density at some reference time. At the level of background cosmology this family
of cosmologies are all equivalent to a standard EdS cosmology driven by a total ``matter'' density $\rho_{tot}=\kappa^2\rho_{m}$. 
However, as soon as perturbations to uniformity are considered, the dependence on the evolution of the scale 
factor on $\kappa^2$ when expressed in terms of the mean (clustering) mass density translates into a physical difference of the models as a function of $\kappa^2$. 
This is seen already, as we will show below, in the evolution of perturbations at linear order. As in the standard EdS case ($\kappa=1$) there is a growing mode and a decaying mode, but their temporal 
evolution is modified when $\kappa\neq 1$. In particular the growing mode for density perturbations has the behaviour $D(a) \propto a^{\alpha}$ 
where
\begin{equation}
\alpha=-\frac{1}{4}+\frac{1}{4}\sqrt{1+\frac{24} {\kappa^2}}\,.\label{eq: alpha}
\end{equation}
As $\alpha$ is a one-to-one function of $\kappa^2$, either can be used to characterize the one-parameter family of 
gEdS models. Because of the crucial role played in perturbation theory by the linear theory growing mode, 
we will often find it convenient to give our results as functions of $\alpha$ in our analysis below. 

We note that while it is required that $\rho_{tot}>0$, and $\rho_m >0$, it is not necessary that $\rho_s>0$. Thus we consider 
the family of gEdS cosmologies to be parameterized either by $\kappa \in [0,\infty]$ or, alternatively, by $\alpha \in [0,\infty]$. 
Our motivation for studying these models here comes, as has been outlined in the introduction, from their interest as a theoretical tool rather than as relevant physical models. We note however that, for $\rho_s>0$ (i.e. $0<\Omega_m<1$), a smooth component with zero pressure can model approximately in certain regimes a component in massive neutrinos, or also a contribution from the zero mode of a scalar field oscillating about the minimum of a quadratic potential or rolling in a simple exponential potential  (see e.g. \cite{ferreira1998cosmology} and references therein). 
An alternative physical interpretation, which 
extends also to models with $\rho_s<0$ (or $\kappa^2<1$), is in terms of a model in which 
the gravitational coupling is scale-dependent, with the expansion being driven by a rescaled
coupling $\kappa^2 G$. The limit
$\kappa \rightarrow 0$ corresponds formally
to the case of a static universe, in which 
a smooth component with negative energy exactly balances the clustering matter
(for further detail, see \cite{benhaiem2014self}). 

\subsection{Eulerian Perturbation Theory (EPT)}

Our starting point is the usual fluid equations for the evolution, under gravity only, of perturbations in the matter in an FLRW universe  in the Newtonian (sub-horizon, non-relativistic) limit (see e.g. 
\cite{Bernardeau2002}):
\begin{widetext}
\begin{eqnarray}
    \frac{\partial \delta (\mathbf{x}, \tau)}{\partial \tau}+\nabla\cdot \big\{[1+\delta (\mathbf{x}, \tau)] \mathbf{u} (\mathbf{x}, \tau)\big\}&=&0\label{eq:continuity},\\
   \frac{\partial \mathbf{u}  (\mathbf{x}, \tau)}{\partial \tau}+\mathcal{H} (\tau) \mathbf{u} (\mathbf{x}, \tau)+[\mathbf{u} (\mathbf{x}, \tau)\cdot\nabla]\mathbf{u} (\mathbf{x}, \tau)
   &=&-\nabla\phi (\mathbf{x}, \tau)-\frac{1}{\rho}\nabla(\rho\sigma_{ij}),\label{eq:euler}\\
   \label{potential_kappa_cosmo}
        \nabla^{2} \phi(\textbf{x},\tau)&=&\frac{3}{2}\Omega_{m}\mathcal{H}^{2}\delta,
    \end{eqnarray}
    \end{widetext}
where $\delta (\mathbf{x}, \tau)\equiv\rho(x,\tau)/\Bar{\rho}(\tau)-1$ is the matter density fluctuation, with $\rho(x, \tau) $ the matter density and $\Bar{\rho}(\tau)$ its mean value, $\mathbf{u}$ is the (peculiar) velocity, $\phi (\mathbf{x}, \tau)$ is the (Newtonian) gravitational potential, and $\sigma_{ij}$ is the velocity dispersion tensor. We will henceforth 
neglect this last term, treating the
fluid in the single flow approximation.

Using the Fourier transform conventions
\begin{equation}
    \begin{split}
        f (\textbf{k})&=\int d^3 r \exp\big[-i\textbf{k}\cdot \textbf{r}\big] f (\textbf{r}),\\
        f (\textbf{r})&=\int \frac{d^3 k}{(2\pi)^{3}}  \exp\big[i\textbf{k}\cdot\textbf{r}\big] f (\textbf{k}),
    \end{split}
\end{equation}
 Eq.~(\ref{eq:continuity}) and the divergence of Eq.~(\ref{eq:euler}) 
 can be combined to give 
\begin{equation}
    \delta'(\textit{\textbf{k}},\tau)+\theta(\textit{\textbf{k}},\tau)=S_{\Tilde{\alpha}}(\textit{\textbf{k}},\tau)\label{eq:density_fourier},
   \end{equation} 
\begin{equation}
    \theta'(\textit{\textbf{k}},\tau)+\mathcal{H}\theta(\textit{\textbf{k}},\tau)+\frac{3}{2}\Omega_{m}(a)\mathcal{H}^{2}\delta(\textit{\textbf{k}},\tau)=S_{\Tilde{\beta}}(\textit{\textbf{k}},\tau)\label{eq:euler_fourier},
    \end{equation} 
where  $\theta$ is the divergence of the velocity field $\mathbf{u}$, so that 
\begin{equation}
    \textbf{u}(\textbf{k})=-i\frac{\textbf{k}}{k^2}\theta(\textbf{k}).
\end{equation}
The velocity field can be assumed to have zero vorticity because the latter can be shown to always decay compared to the irrotational part (see e.g. \cite{Bernardeau2002}). The source terms $S_{\alpha}$ and $S_{\beta}$ in Eqs.~(\ref{eq:density_fourier}) and (\ref{eq:euler_fourier}) are given by
\begin{equation}
    S_{\Tilde{\alpha}}(\textit{\textbf{k}},\tau)=-\int \frac{\text{d}^{3}q}{(2\pi)^{3}}\Tilde{\alpha}(\textit{\textbf{q}},\textit{\textbf{k}}-\textit{\textbf{q}})\theta(\textit{\textbf{q}},\tau)\delta(\textit{\textbf{k}}-\textit{\textbf{q}},\tau)\label{eq:6},
   \end{equation} 
\begin{equation}
    S_{\Tilde{\beta}}(\textit{\textbf{k}},\tau)=-\int \frac{\text{d}^{3}q}{(2\pi)^{3}}\Tilde{\beta}(\textit{\textbf{q}},\textit{\textbf{k}}-\textit{\textbf{q}})\theta(\textit{\textbf{q}},\tau)\theta(\textit{\textbf{k}}-\textit{\textbf{q}},\tau)\label{eq:7},
   \end{equation} 
   where the coupling kernels $\Tilde{\alpha}$ and $\Tilde{\beta}$ respectively are defined as
   
 \begin{equation}
   \Tilde{\alpha}(\textit{\textbf{q}}_{1},\textit{\textbf{q}}_{2})=\frac{\textit{\textbf{q}}_{1}.(\textit{\textbf{q}}_{1}+\textit{\textbf{q}}_{2})}{\textit{q}_{1}^{2}}\label{eq:8},
  \end{equation} 
   \begin{equation}
   \Tilde{\beta}(\textit{\textbf{q}}_{1},\textit{\textbf{q}}_{2})=\frac{1}{2}(\textit{\textbf{q}}_{1}+\textit{\textbf{q}}_{2})^{2}\frac{\textit{\textbf{q}}_{1}.\textit{\textbf{q}}_{2}}{\textit{q}_{1}^{2}\textit{q}_{2}^{2}}\label{eq:9}.
  \end{equation} 
For a gEdS cosmology, parametrized by 
$\kappa^2$, combining Eqs.~(\ref{eq:density_fourier}) and (\ref{eq:euler_fourier}) with Eq.~(\ref{potential_kappa_cosmo})
to obtain a single second-order differential equation for the density fluctuations $\delta (\textbf{x},\tau)$ and for the divergence of velocity fluctuations  $\theta(\textbf{x},\tau)$, respectively, we have
\begin{widetext}
\begin{eqnarray}
    \mathcal{H}^{2}\Big\{-a^{2}\partial_{a}^{2}-\frac{3}{2}a\partial_{a}+\frac{3}{2\kappa^{2}}\Big\}\delta(\textit{\textbf{k}},a)&=&S_{\Tilde{\beta}}(\textit{\textbf{k}},a)-\mathcal{H}\partial_{a}(a S_{\Tilde{\alpha}}(\textit{\textbf{k}},a))\label{equation-second-order-density},\\
    \mathcal{H}\Big\{a^{2}\partial_{a}^{2}+\frac{5}{2}a\partial_{a}+\Big(\frac{1}{2}-\frac{3}{2\kappa^{2}}\Big)\Big\}\theta(\textit{\textbf{k}},a)&=&\partial_{a}(a S_{\Tilde{\beta}}(\textit{\textbf{k}},a))-\frac{3}{2\kappa^{2}}\mathcal{H}S_{\Tilde{\alpha}}(\textit{\textbf{k}},a)\label{eq:11},
   \end{eqnarray} 
\end{widetext}
where $\partial_{a}$ is the partial derivative respect to scale factor $a$. 

We now proceed to solve these equations following the standard method (see e.g. \cite{Bernardeau2002}) in standard
Eulerian perturbation theory (EPT). Working to linear order in $\delta$ and $\theta$, Eq.~(\ref{equation-second-order-density})
is just
\begin{equation}
   -a^{2}\partial_{a}^{2}\delta(\textit{\textbf{k}},a)-\frac{3}{2}a\partial_{a}\delta(\textit{\textbf{k}},a)+\frac{3}{2\kappa^{2}}\delta(\textit{\textbf{k}},a)=0,
   \end{equation} 
which has linearly independent decaying and growing mode solutions so that
\begin{equation}
\begin{split}
  \delta(\textit{\textbf{k}},a)&=\delta_{+}(\textbf{k},a)+\delta_{-}(\textbf{k},a)\\
  &=D_{+}(a)\delta_{+,0}(\textbf{k})+D_{-}(a)\delta_{-,0}(\textbf{k}),
\end{split}
\end{equation}
where $\delta_{\pm,0}(\textbf{k})$ are constants fixed at some reference time, $a=1$, and $D_{\pm}=a^{\alpha_\pm}$ with 
\begin{equation}
    \alpha_\pm=-\frac{1}{4}\pm\frac{1}{4}\sqrt{1+\frac{24}{\kappa^{2}}}.\label{eq: 2.13}
\end{equation}

We assume the growing mode initial conditions, corresponding to 
the solution at linear order which may be written
\begin{equation}
\label{delta-linear}
    \delta(\textit{\textbf{k}},a)=D (a)\delta^{(1)} (\textbf{k}),
\end{equation}
where $D (a)=a^\alpha$, and $\delta^{(1)} (\textbf{k})$ is the fluctuation at $a=1$.
At the same linear order we have the relation between density fluctuations and velocity 
\begin{equation}\label{theta_linear_solution}
    \theta(\textit{\textbf{k}},\tau)=-\delta'(\textit{\textbf{k}},\tau)\,,
   \end{equation} 
from which the solution at linear order for the velocity divergence
follows, 
\begin{equation}
\label{vel-linear}
    \theta(\textit{\textbf{k}},a)=-\mathcal{H} (a) \alpha D(a) \delta^{(1)} (\textbf{k}).
   \end{equation} 
 For the case $\kappa=1$ (and $\alpha=1$) we recover the standard expression for the EdS model. These expressions are just special cases of the standard ones for FLRW cosmologies,
 which correspond to Eq.~(\ref{delta-linear}) and 
 \begin{equation} 
 \label{vel-linear-general}
    \theta(\textit{\textbf{k}},a)=-\mathcal{H} (a) \alpha_1(a) D_1(a) \delta^{(1)} (\textbf{k}),
   \end{equation} 
where $D_1(a)$ is the appropriate growth factor for the linear theory growing mode in the cosmology, and 
\begin{equation}
\alpha_1(a)=\frac{d\ln D_1}{d\ln a}. 
\end{equation}
 
 Taking leading order solutions as Eqs.~(\ref{delta-linear}) and (\ref{vel-linear}), we now proceed to solve Eqs.~(\ref{equation-second-order-density}) and (\ref{eq:11}) perturbatively to higher orders taking the ansatz 
\begin{equation}
\label{separable_ansatz}
    \begin{split}
        \delta(\textbf{k},a)&=\sum_{i=1}^{\infty}D^{i}(a)\,\delta^{(i)}(\textbf{k}), \\
        \theta(\textbf{k},a)&=-\mathcal{H} (a) \alpha \sum_{i=1}^{\infty}D^{i}(a)\,\theta^{(i)}(\textbf{k}),
        \end{split}
\end{equation}
where $D^i (a)\equiv (a^\alpha)^i$, and $\delta^{(i)}$, $\theta^{(i)}(\textbf{k})$ are the contributions at order $i$ to
each of these quantities. Such a ``separable'' ansatz means that all dependence on the cosmological model disappears when $\delta(\textbf{k})$ and $\theta (\textbf{k})$ are expressed in terms of their values
at linear order. Such an ansatz solves the equations exactly in perturbation for the EdS model (see e.g. \cite{Bernardeau2002}). For the family of perturbed FLRW cosmologies with a perturbed matter component, one can consider the same ansatz with $\alpha$ replaced by 
$\alpha_1(a)$, and it has been shown \cite{martel371second,scoccimarro1998nonlinear}
that the condition for it be a solution  
is that the ratio
\begin{equation}
\frac{\Omega_m}{\alpha_1^2},
\end{equation}
is constant in time. While this condition is not satisfied in LCDM-like models, it is  in gEdS models (in which both $\Omega_m$ and $\alpha_1$ are individually 
constant). 

Using this ansatz (\ref{separable_ansatz}) 
the \textit{n}-th order solutions for the density and velocity fields can be 
expressed in exactly the same form as in standard EdS as 
\begin{eqnarray}\label{ansatz-delta-theta}
    \delta^{(n)}(\textbf{k})&=&\prod_{m=1}^{n}\Big\{\int\frac{d^{3}q_{m}}{(2\pi)^{3}}\delta^{(1)}(\textbf{q}_{m})\Big\}\nonumber\\ 
    & &\times (2\pi)^{3}\delta_{D}(\textbf{k}-\textbf{q}|_{1}^{n}) F_{n}(\textbf{q}_{1},\dots,\textbf{q}_{n}),\\
\theta^{(n)}(\textbf{k})&=&\prod_{m=1}^{n}\Big\{\int\frac{d^{3}q_{m}}{(2\pi)^{3}}\delta^{(1)}(\textbf{q}_{m})\Big\}\nonumber \\
& &\times (2\pi)^{3}\delta_{D}(\textbf{k}-\textbf{q}|_{1}^{n})
 G_{n}(\textbf{q}_{1},\dots,\textbf{q}_{n}).  
\end{eqnarray}
Working to second-order in the 
expansion, we have
\begin{eqnarray}\label{derivation-second-order-density}
     & &\mathcal{H}^{2}\Big\{-a^{2}\partial_{a}^{2}-\frac{3}{2}a\partial_{a}+\frac{3}{2\kappa^{2}}\Big\}a^{2\alpha}\delta^{(2)}(\textit{\textbf{k}})\nonumber\\
     & =&-\int \frac{\text{d}^{3}q}{(2\pi)^{3}}\Tilde{\beta}(\textit{\textbf{q}},\textit{\textbf{k}}-\textit{\textbf{q}})\delta^{(1)}(\textit{\textbf{q}})\delta^{(1)}(\textit{\textbf{k}}-\textit{\textbf{q}})
     \times\big[\mathcal{H}^{2}\alpha^{2}a^{2\alpha}\big]\nonumber\\
     & &-\int \frac{\text{d}^{3}q}{(2\pi)^{3}}\Tilde{\alpha}(\textit{\textbf{q}},\textit{\textbf{k}}-\textit{\textbf{q}})\delta^{(1)}(\textit{\textbf{q}})\times\delta^{(1)}(\textit{\textbf{k}}-\textit{\textbf{q}})\nonumber\\
     & &\times\mathcal{H}^{2}\big[\frac{\alpha}{2}+2\alpha^2\big]a^{2\alpha},
   \end{eqnarray} 
from which we obtain the second-order density kernel $\textit{F}_{2}(\textit{\textbf{q}}_{1},\textit{\textbf{q}}_{2})$ 
as 
\begin{equation}\label{densityKernel_0}
      \Big(3\alpha^2+\frac{\alpha}{2}\Big)\textit{F}_{2}(\textit{\textbf{q}}_{1},\textit{\textbf{q}}_{2})=\alpha^{2}\Tilde{\beta}(\textit{\textbf{q}}_{1},\textit{\textbf{q}}_{2})+ \Big(\frac{\alpha}{2}+2\alpha^2\Big)\Tilde{\alpha}(\textit{\textbf{q}}_{1},\textit{\textbf{q}}_{2}), \nonumber
  \end{equation} 
and thus finally 
\begin{equation}\label{densityKernel}
       \textit{F}_{2}(\textit{\textbf{q}}_{1},\textit{\textbf{q}}_{2})=d_{2} \Tilde{\alpha}(\textit{\textbf{q}}_{1},\textit{\textbf{q}}_{2})+(1-d_{2}) \Tilde{\beta}(\textit{\textbf{q}}_{1},\textit{\textbf{q}}_{2}),
  \end{equation} 
 where
  \begin{equation}
      d_{2}=\Big(\frac{1+4\alpha}{1+6\alpha}\Big).
  \end{equation}
For the velocity field divergence 
at the same order we have
   \begin{eqnarray}\label{derivation-second-order-velocity}
        & &\mathcal{H}\Big\{a^{2}\partial_{a}^{2}+\frac{5}{2}a\partial_{a}+(\frac{1}{2}-\frac{3}{2\kappa^{2}})\Big\}\big[-\mathcal{H}\alpha a^{2\alpha}\theta^{(2)}(\textit{\textbf{k}})\big]\nonumber\\
        &=&-\int \frac{\text{d}^{3}q}{(2\pi)^{3}}\Tilde{\beta}(\textit{\textbf{q}},\textit{\textbf{k}}-\textit{\textbf{q}})\delta^{(1)}(\textit{\textbf{q}})\delta^{(1)}(\textit{\textbf{k}}-\textit{\textbf{q}})\nonumber\\
        & &\times\big[\mathcal{H}^{2}2\alpha^{3}a^{2\alpha}\big]-\int \frac{\text{d}^{3}q}{(2\pi)^{3}}\Tilde{\alpha}(\textit{\textbf{q}},\textit{\textbf{k}}-\textit{\textbf{q}})\nonumber\\
        & &\times\delta^{(1)}(\textit{\textbf{q}})\delta^{(1)}(\textit{\textbf{k}}-\textit{\textbf{q}})\mathcal{H}^{2}\big[\frac{\alpha^2}{2}+\alpha^3\big]a^{2\alpha},
   \end{eqnarray}
which leads to the solution
\begin{equation}
       \textit{G}_{2}(\textit{\textbf{q}}_{1},\textit{\textbf{q}}_{2})=\tilde{d}_{2} \Tilde{\alpha}(\textit{\textbf{q}}_{1},\textit{\textbf{q}}_{2})+(1-\tilde{d}_{2}) \Tilde{\beta}(\textit{\textbf{q}}_{1},\textit{\textbf{q}}_{2})
      \end{equation}
where 
  \begin{equation}
      \tilde{d}_{2}=\Big(\frac{1+2\alpha}{1+6\alpha}\Big).
  \end{equation}
   
Continuing to third-order in the same manner
we obtain 
\begin{eqnarray}\label{kernel_F3}
      \textit{F}_{3} (\textit{\textbf{q}}_{1},\textit{\textbf{q}}_{2},\textit{\textbf{q}}_{3}) &=&\frac{1}{2}\bigg[d_{3}\Tilde{\alpha}(\textit{\textbf{q}}_{1},\textit{\textbf{q}}_{2}+\textit{\textbf{q}}_{3})\textit{F}_{2}(\textit{\textbf{q}}_{2},\textit{\textbf{q}}_{3})\nonumber\\& &+(1-d_{3})\Tilde{\beta}(\textit{\textbf{q}}_{1},\textit{\textbf{q}}_{2}+\textit{\textbf{q}}_{3})\textit{G}_{2}(\textit{\textbf{q}}_{2},\textit{\textbf{q}}_{3})\nonumber \\
      & &+\big[d_{3}\Tilde{\alpha}(\textit{\textbf{q}}_{1}+\textit{\textbf{q}}_{2},\textit{\textbf{q}}_{3})\nonumber\\
      & &+(1-d_{3})\Tilde{\beta}(\textit{\textbf{q}}_{1}+\textit{\textbf{q}}_{2},\textit{\textbf{q}}_{3})\big]\nonumber\\
      & & \times \textit{G}_{2}(\textit{\textbf{q}}_{1},\textit{\textbf{q}}_{2}) \bigg],
 \end{eqnarray}
and 
\begin{eqnarray}\label{kernel_G3}
      \textit{G}_{3} (\textit{\textbf{q}}_{1},\textit{\textbf{q}}_{2},\textit{\textbf{q}}_{3})& =&\frac{1}{2}\bigg[\tilde{d}_{3}\Tilde{\alpha}(\textit{\textbf{q}}_{1},\textit{\textbf{q}}_{2}+\textit{\textbf{q}}_{3})\textit{F}_{2}(\textit{\textbf{q}}_{2},\textit{\textbf{q}}_{3})\nonumber \\& & +(1-\tilde{d}_{3})\Tilde{\beta}(\textit{\textbf{q}}_{1},\textit{\textbf{q}}_{2}+\textit{\textbf{q}}_{3})\textit{G}_{2}(\textit{\textbf{q}}_{2},\textit{\textbf{q}}_{3})\nonumber\\
      & &+\big[\tilde{d}_{3}\Tilde{\alpha}(\textit{\textbf{q}}_{1}+\textit{\textbf{q}}_{2},\textit{\textbf{q}}_{3})\nonumber \\
      & &+(1-\tilde{d}_{3})\Tilde{\beta}(\textit{\textbf{q}}_{1}+\textit{\textbf{q}}_{2},\textit{\textbf{q}}_{3})\big]\nonumber\\
      & &\times\textit{G}_{2}(\textit{\textbf{q}}_{1},\textit{\textbf{q}}_{2}) \bigg],
   \end{eqnarray}
 where
  \begin{equation}
      d_{3}=\frac{1+6\alpha}{1+8\alpha}, \quad \tilde{d}_{3}=\frac{1+2\alpha}{1+8\alpha}.
  \end{equation}  
Higher-order kernels can be inferred by continuing this recursion.\footnote{Further details of these calculations are given in Appendix \ref{third_order_kernels_derivation}.}

\begin{figure}[t]
\includegraphics[width=7.5cm, height=7.5cm]{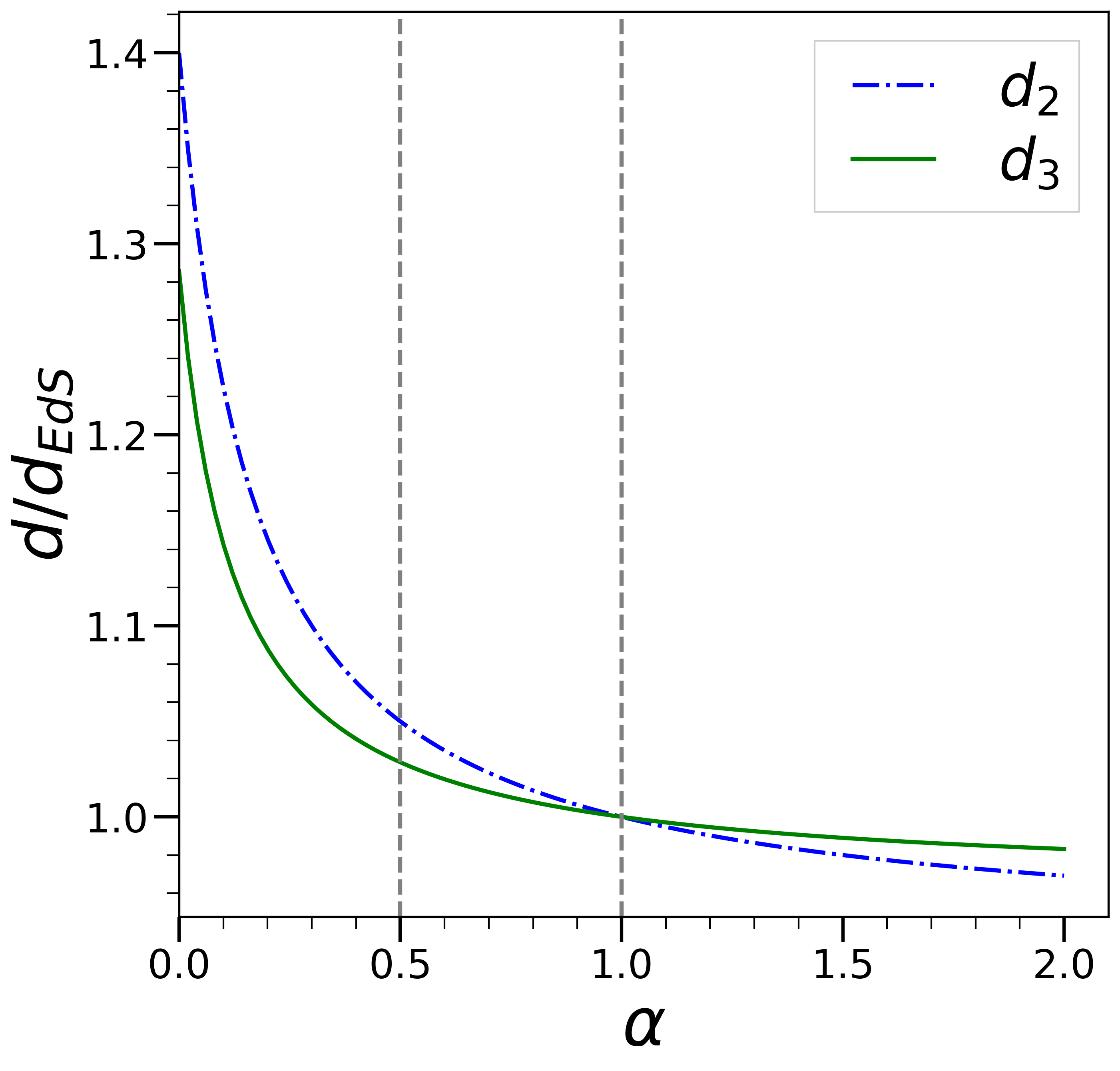}
\caption{Variation of the functions $d_2$ and $d_3$ determining the EPT kernels at second and third order relative to their values in EdS, as a function of the logarithmic growth rate $\alpha$ 
of gEdS. The dashed horizontal lines bracket approximately the region in which the effective growth rate $\alpha_1(a)$ varies in LCDM-like models.}
\label{d_2_d_3vs_alpha}
\end{figure}

  Fig.~\ref{d_2_d_3vs_alpha} shows how the kernels $F_2$ and $G_2$ vary relative to those in the standard EdS case as a function of $\alpha$ (and $F_3$ and $G_3$ have very similar behaviours).
  We see that the dependence on $\alpha$ is very weak, other than at values of $\alpha$ very much less than unity. Indeed expanding about
  $\alpha=1$ we have 
  \begin{eqnarray}
  \label{kernels_alpha_expansion}
      d_2=d_2(1)\left[1+\frac{2}{35} (1-\alpha) + O((1-\alpha)^2) \right], \nonumber \\
      d_3=d_3(1)\left[1+\frac{2}{63} (1-\alpha) + O((1-\alpha)^2) \right],
  \end{eqnarray}
  and, over the whole range of $\alpha$, varying from $0$ to $\infty$, we have
  \begin{eqnarray}
      \frac{7}{5} > \frac{d_2}{d_2(1)} > \frac{14}{15}, \\
      \frac{9}{7} >  \frac{d_3}{d_3(1)} > \frac{27}{28}.
  \end{eqnarray}
A naive guess would be that, in a generic FLRW cosmology, the kernel at any time can be approximated by those in a gEdS model, with the growth rate corresponding to the instantaneous value of its effective logarithmic growth rate 
$\alpha_{1}(a)$. For standard-type cosmological models, as we will discuss in detail in the next section, such an effective growth rate decreases relative to EdS to at most $\alpha\approx 0.5$ at $z=0$ as dark energy contributions accelerate the expansion. From Fig.~\ref{d_2_d_3vs_alpha} we see that this would correspond to a change
of about $5\%$ ($3\%$) in $d_2$($d_3$) relative to EdS. In practice, as we will see in the second part of the paper, such a mapping to an approximate mapping
may indeed to used, but 
the relevant effective growth exponent is given by this naive guess only for a sufficiently slowly varying smooth component. For LCDM-like models it is much closer to the
initial value. The very small sub-percent corrections calculated for LCDM-like models can thus be understood as a combination of the very weak sensitivity of the gEdS kernels to the growth rate of fluctuations, and the rapid evolution of dark energy at low redshift. 

\subsection{Lagrangian Perturbation Theory (LPT)}
In LPT one characterizes the evolution of the fluid using its displacement field $\Psi(\textbf{q},t)$ as a function of its initial Lagrangian position $\textbf{q}$ of the particle, in terms of which the Eulerian position is given by  
\begin{equation}
    \textbf{x}(\textbf{q},\tau)=\textbf{q}+\mathbf{\Psi}(\textbf{q},\tau).
\end{equation}
It follows in particular 
that 
\begin{equation}
    \big[1+\delta(\textbf{x})\big]\text{d}^{3} \text{x}=\text{d}^{3} \text{q}\label{eq:20},
\end{equation}
and that the determinant $J$ of the 
Jacobian matrix is given by 
\begin{equation}
    J=\Big|\frac{d^{3}x}{d^{3}q}\Big|=\text{det}\Big[\delta_{ij}^{(K)}+\Psi_{i,j}\Big]\label{eq:21},
\end{equation}
where
\begin{equation}
    \Psi_{i,j}=\frac{\partial\Psi_{i}}{\partial q_{j}},
\end{equation}
so that 
\begin{equation}
    \delta(\textbf{x})=\frac{1}{J}-1.
\end{equation}
The equation of motion for the 
fluid particle is 
\begin{equation}
    \frac{\text{d}^{2}\mathbf{\Psi}}{\text{d}\tau^{2}}+\mathcal{H}\frac{\text{d}\mathbf{\Psi}}{\text{d}\tau}=-\nabla_{x}\phi\label{eq:24}.
\end{equation}
Taking the divergence of this equation, and using the Poisson Eq. (\ref{potential_kappa_cosmo}), one obtains
\begin{equation}
    \begin{split}
             J \boldsymbol{\nabla}_{\textbf{x}}\cdot\Big[\frac{\text{d}^{2}\mathbf{\Psi}}{\text{d}\tau^{2}}+\mathcal{H}\frac{\text{d}\mathbf{\Psi}}{\text{d}\tau}\Big]&=\frac{3}{2}\Omega_{m}\mathcal{H}^{2}[J-1]\label{eq:25}.
    \end{split}
\end{equation}
This equation order is then solved order by order using the perturbative expansion 
for 
\begin{equation}
    \mathbf{\Psi}=\mathbf{\Psi}^{(1)}+\mathbf{\Psi}^{(2)}+\mathbf{\Psi}^{(3)}+\dots\label{eq:27}
\end{equation}
and
\begin{equation}
    J=1+J^{(1)}+J^{(2)}+J^{(3)}+\dots\label{eq:28}
\end{equation}
where the latter are related to the former up to third order by the 
following expressions (for a detailed derivation, see e.g. \cite{bouchet1995perturbative}):
\begin{equation}
    \begin{split}
        J^{(1)}&=\sum_{i}\Psi_{i,i}^{(1)},\\
        J^{(2)}&=\sum_{i}\Psi_{i,i}^{(2)}+\frac{1}{2}\sum_{i\neq j}\Big\{\Psi_{i,i}^{(1)}\Psi_{j,j}^{(1)}-\Psi_{i,j}^{(1)}\Psi_{j,i}^{(1)}\Big\},\\
        J^{(3)}&=\sum_{i}\Psi_{i,i}^{(3)}+\sum_{i\neq j}\Big\{\Psi_{i,i}^{(2)}\Psi_{j,j}^{(1)}-\Psi_{i,j}^{(2)}\Psi_{j,i}^{(1)}\Big\}+\text{det}\Psi_{i,j}^{(1)}\label{eq:29}.
    \end{split}
\end{equation}
Using the chain rule 
\begin{equation}
    \nabla_{x_{i}}=(\delta_{i j}^{(K)}+\Psi_{i,j})^{-1}\nabla_{q_{j}},
\end{equation}
one then solves Eq.~(\ref{eq:25}) 
order by order.
At linear order we have simply
\begin{equation}
   \frac{\text{d}^{2}\Psi_{i,i}^{(1)}}{\text{d}\tau^{2}}+\mathcal{H}\frac{\text{d}\Psi_{i,i}^{(1)}}{\text{d}\tau}-\frac{3}{2}\Omega_{m}\mathcal{H}^{2}\Psi_{i,i}^{(1)}=0,
\end{equation}
and thus, choosing the growing mode solution,  
\begin{equation}
    \Psi_{i,i}^{(1)}(\textbf{x},\tau)=-\delta^{(1)}(\textbf{x},\tau),
\end{equation}
which gives in Fourier space
\begin{equation}
    \Psi^{(1)}(\textbf{k},\tau)=-\frac{i\textbf{k}}{k^{2}}\delta^{(1)}(\textbf{k})D_1(\tau).
\end{equation}
This is the solution for any perturbed FLRW cosmology and thus for gEdS in particular with $D_1(a)=D(a)=a^\alpha$. 

Considering now gEdS cosmologies, we proceed to solve Eq.~(\ref{eq:25})  to non-linear order making the separable ansatz analogous to that in EPT:
\begin{equation}
        \Psi^{(n)}(\textbf{k},\tau)=D^{n}(\tau) \Psi^{(n)}(\textbf{k})\,.
\end{equation}
We have then
\begin{equation}
\begin{split}
  &(1+J^{(1)}+J^{(2)}+J^{(3)}+\dots)(\delta_{i j}^{(K)}-\Psi_{i,j}+\Psi_{i,l}\Psi_{l,j}+\dots)\\
    &\times n\alpha \Big[n \alpha+\frac{1}{2}\Big]\Psi_{i,j}=\frac{3}{2\kappa^{2}}[J^{(1)}+J^{(2)}+J^{(3)}+\dots],\label{equation_displacement}
\end{split}
\end{equation}
where we have used 
$\frac{\text{d}\mathcal{H}}{\text{d}a}=-\frac{\mathcal{H}}{2a}$.
At second order we have therefore 
\begin{equation}
        \Psi_{i,i}^{(2)}=-\frac{2\alpha+1}{2(6\alpha+1)}\sum_{i,j}\big[\Psi_{i,i}^{(1)}\Psi_{j,j}^{(1)}-\Psi_{i,j}^{(1)}\Psi_{i,j}^{(1)}\Big]\label{eq:42}.
\end{equation}
The result at third-order, of which further details are given in  Appendix \ref{Appendix-Lagrangian perturbation theory}, is 
\begin{eqnarray}
      \Psi_{i,i}^{(3)}&=&\frac{1}{(8\alpha+1)}\Big[(4\alpha+1)\Big\{-\Psi_{i,i}^{(2)}\Psi_{j,j}^{(1)}+\Psi_{i,j}^{(2)}\Psi_{j,i}^{(1)}\Big\}\nonumber\\
      & &+(2\alpha+1)\Big\{\frac{1}{2}\Psi_{i,i}^{(1)}\Psi_{i,j}^{(1)}\Psi_{j,i}^{(1)}\nonumber \\
       & &-\frac{1}{6}\Psi_{i,i}^{(1)}\Psi_{j,j}^{(1)}\Psi_{k,k}^{(1)}-\frac{1}{3}\Psi_{i,k}^{(1)}\Psi_{k,j}^{(1)}\Psi_{j,i}^{(1)}\Big\}\Big].\label{Third-order-LPT}
\end{eqnarray}
At \textit{n}-th order the displacement field in Fourier space can be written in the form
\begin{eqnarray}
  \Psi^{(n)}(\textbf{k})&=&-\frac{i}{n}\prod_{i=1}^{n}\Big\{\int \frac{\text{d}^{3}q_{i}}{(2\pi)^{3}}\delta^{(1)}(\textbf{q}_{i})\Big\}\textbf{L}_{n}(\textbf{q}_{1},\dots,\textbf{q}_{n})\nonumber \\
  & &\times (2\pi)^{3}\delta^{(D)}(\textbf{k}-\textbf{q}_{1}\dots\textbf{q}_{n}),
\end{eqnarray}
where the first, second, and third-order solutions correspond to the following  kernels:
\begin{eqnarray}
         \textbf{L}_{1}&=&\frac{\textbf{k}}{k^{2}},\\
         \textbf{L}_{2}&=&\Big(\frac{2\alpha+1}{6\alpha+1}\Big)\frac{\textbf{k}}{k^{2}}\Big[1-\frac{(\textbf{q}_{1}\cdot\textbf{q}_{2})^{2}}{q_{1}^{2}q_{2}^{2}}\Big],\\
         \textbf{L}_{3}&=&3\Big(\frac{4\alpha+1}{8\alpha+1}\Big)\Big(\frac{2\alpha+1}{6\alpha+1}\Big)\frac{\textbf{k}}{k^{2}}\Big[1-\Big(\frac{\textbf{q}_{1}\cdot\textbf{q}_{2}}{q_{1}q_{2}}\Big)^{2}\Big]\nonumber \\
         & &\times\Big[1-\Big(\frac{(\textbf{q}_{1}+\textbf{q}_{2})\cdot\textbf{q}_{3}}{|(\textbf{q}_{1}+\textbf{q}_{2}|q_{3}}\Big)^{2}\Big]-\Big(\frac{2\alpha+1}{8\alpha+1}\Big)\frac{\textbf{k}}{k^{2}}\nonumber\\
         & &\times\Big[1-3\Big(\frac{\textbf{q}_{1}\cdot\textbf{q}_{2}}{q_{1}q_{2}}\Big)^{2}\nonumber\\
         & &+2\frac{(\textbf{q}_{1}\cdot\textbf{q}_{2})(\textbf{q}_{2}\cdot\textbf{q}_{3})(\textbf{q}_{3}\cdot\textbf{q}_{1})}{q_{1}^{2}q_{2}^{2}q_{3}^{2}}\Big].
\end{eqnarray}
As a check on our calculation we can use the known relations between the EPT and LPT kernels (see \cite{Aviles2018Nonlinear, Matsubara, Rampf}). At first-order we have simply
\begin{equation}
    F_{1}(\textbf{k})=\textbf{k}\cdot\textbf{L}_{1}(\textbf{k})=\textbf{k}\cdot\frac{\textbf{k}}{k^{2}}=1.
\end{equation}
The second-order density kernels are related by
\begin{eqnarray}
        F_{2}(\textbf{q}_{1},\textbf{q}_{2})
        &=&\frac{1}{2}\Big[\textbf{k}\cdot\textbf{L}_{2}(\textbf{q}_{1},\textbf{q}_{2})+[\textbf{k}\cdot\textbf{L}_{1}(\textbf{q}_{1}][\textbf{k}\cdot\textbf{L}_{1}(\textbf{q}_{2})]\Big]\nonumber\\
        &=&\frac{4\alpha+1}{6\alpha+1}+\frac{1}{2}\frac{\textit{\textbf{q}}_{1}\cdot\textit{\textbf{q}}_{2}}{q_{1}q_{2}}\Big[\frac{q_{2}}{q_{1}}+\frac{q_{1}}{q_{2}}\Big]\nonumber\\
        & &+\Big(\frac{2\alpha}{6\alpha+1}\Big)\frac{(\textit{\textbf{q}}_{1}\cdot\textit{\textbf{q}}_{2})^{2}}{q_{1}^{2}q_{2}^{2}},
\end{eqnarray}
while at third order we have
\begin{eqnarray}
        F_{3}(\textbf{q}_{1},\textbf{q}_{2},\textbf{q}_{3})
        &=&\frac{1}{3!}\Big[\textbf{k}\cdot\textbf{L}_{3}^{(s)}(\textbf{q}_{1},\textbf{q}_{2},\textbf{q}_{3})\nonumber \\
        & &+\big\{[\textbf{k}\cdot\textbf{L}_{1}(\textbf{q}_{1})][\textbf{k}\cdot\textbf{L}_{2}(\textbf{q}_{2},\textbf{q}_{3})]\nonumber \\ & &+\text{cyc}\big\}\nonumber\\
        & &+[\textbf{k}\cdot\textbf{L}_{1}(\textbf{q}_{1})][\textbf{k}\cdot\textbf{L}_{1}(\textbf{q}_{2})]\nonumber \\ & &\times [\textbf{k}\cdot\textbf{L}_{1}(\textbf{q}_{3})]\Big].
\end{eqnarray}
It is straightforward to check that we indeed recover the results in EPT of the previous subsection.

\section{Power spectrum in gEdS models}
\label{PS-gEdS-models}
We define the power spectrum $P(\vec{k}) \equiv P(k)$ ($k=|\textbf{k}|$) of the (assumed) statistically homogeneous and isotropic stochastic density field by  
\begin{equation}\label{definition-PS}
    \langle \delta(\textbf{k},a)\delta(\textbf{k}',a) \rangle=(2\pi)^3 \delta^{(D)} (\textbf{k}+\textbf{k}')P(|\textbf{k}|,a),
\end{equation}
where $\langle \cdots \rangle$ denotes the ensemble average. Assuming that the fluctuations are Gaussian at linear order, 
one obtains, retaining terms up to fourth order in the linear order solution,  
\begin{eqnarray}\label{power_spectrum}
    \langle \delta(\textbf{k},a)\delta(\textbf{k}',a) \rangle&=&\langle \delta^{(1)}(\textbf{k},a)\delta^{(1)}(\textbf{k}',a) \rangle\nonumber \\
    & &+2\langle \delta^{(1)}(\textbf{k},a)\delta^{(3)}(\textbf{k}',a) \rangle\nonumber \\
    & &+\langle\delta^{(2)}(\textbf{k},a)\delta^{(2)}(\textbf{k}',a) \rangle,
\end{eqnarray}
and thus the expression for the ``one-loop'' power spectrum as 
\begin{equation}\label{power_spectrum1}
 \textit{P}_{1-\text{loop}}(k,a)=P_{\text{\rm lin}}(k,a)+P_{22}(k,a)+2P_{13}(k,a),
\end{equation}
where $P_{\text{lin}}(k,a)$ is the linear power spectrum (i.e. defined by
Eq.~(\ref{definition-PS}) with $\delta(\vec{k},a)=D(a)\delta^{(1)}(\vec{k}$), 
and the two other terms correspond to 
 \begin{eqnarray}
     P_{22}(k,a)&=&2\int \frac{d^{3}q}{(2\pi)^3} P_{\text{lin}}(q,a) P_{\text{lin}}(|\textbf{k}-\textbf{q}|,a)\nonumber \\
     & &\times|F_{2}^{(s)}(\textbf{k}-\textbf{q},\textbf{q})|^{2}.\label{equation-P22}
 \end{eqnarray}
and 
\begin{eqnarray}
     P_{13}(k,a)&=&3 P_{\text{lin}}(k,a) \int \frac{d^{3}q}{(2\pi)^3} P_{\text{lin}}(q,a) \nonumber \\
     & &\times F_{3}^{(s)}(\textbf{k},\textbf{q},-\textbf{q})\label{P13-alpha},
 \end{eqnarray}
where $F_{2}^{(s)}$ and $F_{3}^{(s)}$ are the symmetrized form of $F_2$ and $F_3$
(with respect to permutation of their arguments). These expressions are formally 
identical to those in standard EdS, and share the property of any separable solution 
that the perturbative corrections to the PS at any time are functionals only of the 
linear power spectrum at that time.
 The modifications of gEdS relative to the standard EdS model are expressed solely through the $\alpha$-dependence of the kernels $F_2$ and $F_3$ which we have derived above. 
 Using these to make the $\alpha$-dependence explicit we obtain 
\begin{eqnarray} \label{P22-gEdS}
P_{22}&=&M_0+\frac{1+4\alpha}{1+6\alpha} M_1 + \left(\frac{1+4\alpha}{1+6\alpha}\right)^2 M_2, \\
\label{P13-gEdS}
2P_{13}&=&N_0+ \frac{1+2\alpha}{1+8\alpha} N_1 +\frac{2\alpha(1+2\alpha)}{(1+6\alpha)(1+8\alpha)} N_2,
\end{eqnarray}
where the $M_i$ are the integrals
\begin{eqnarray}\label{Equation-P22-M}
M_i (k,a)&=&\frac{1}{8\pi^2}k^{3}\int_{0}^{\infty} d r P_{\text{lin}} (k r,a) \nonumber \\
& &\times \int_{-1}^{1} d \mu \frac{P_{\text{lin}} (k\sqrt{1+r^2 -2\mu r},a)}{(1+r^2 -2\mu r)^2}\,\nonumber\\
& &\times m_i(r,\mu),
\end{eqnarray}
with
\begin{eqnarray}\label{Equation-P22-M-defs}
m_0 (r,\mu)&=&(\mu-r)^2,\\
m_1 (r,\mu)&=& 4r(\mu-r)(1-\mu^2),\\
m_2 (r,\mu)&=& 4r^2 (1-\mu^2)^2,
\end{eqnarray}
and $N_i$ are the integrals
\begin{equation}\label{Equation-P13-N}
N_i(k,a)=\frac{1}{8\pi^2}k^{3} P_{\rm lin} (k,a) \int_{0}^{\infty} dr P_{\rm lin} (kr,a) n_i (r)
\end{equation}
with
\begin{equation}\label{Equation-P13-N-defs}
     \begin{split}
     n_0(r)&=-\frac{4}{3},\\
     n_1(r)&=1+\frac{8}{3}r^2-r^4+\frac{(r^{2}-1)^{3}}{2r} \ln\frac{|1+r|}{|1-r|}, \\
     n_2(r)&=\frac{1}{r^2}(1-\frac{8}{3}r^2-r^4)+\frac{(r^{2}-1)^{3}}{2r^3} \ln\frac{|1+r|}{|1-r|},
         \end{split}
\end{equation}
and we have defined $\mu=\textbf{k}\cdot\textbf{q}/kq$ and $r=q/k$.


\subsection{Convergence properties of the one-loop power spectrum}
\begin{table}[tpb]
\centering
\begin{tabular}{|c c c |} 
 \hline
  & IR-divergent & UV-divergent  \\ [0.5ex] 
 \hline\hline
 $M_{0}$ & $n\leq-1$ & $n\geq1/2$  \\
 $M_{1}$ & $n\leq-3$ & $n\geq1/2$  \\
 $M_{2}$ & $n\leq-3$ & $n\geq1/2$  \\
  \hline\hline
 $N_{0}$ & $n\leq-1$ & $n\geq-1$  \\
 $N_{1}$ & $n\leq-3$& $n\geq-1$\\
 $N_{2}$ & $n\leq-3$ & $n\geq-1$\\
  \hline\hline
 $N_{0}+M_{0}$ & $n\leq-3$ & $n\geq-1$  \\ 
 [1ex] 
 \hline
\end{tabular}
\caption{Convergence properties of the integrals $M_i$ and $N_i$ ($i=0,1,2$). In each case the bound on $n$ is that obtained assuming that
$P_{\rm lin} \sim k^n$ in the relevant  ($k \rightarrow 0$ or $k \rightarrow \infty$) limit. Integrals are ``infra-red safe'' if they converge
for $n > -3$, i.e. when $P_{\rm lin}(k)$  itself is integrable (in three dimensions) as $k \rightarrow 0$. The cancellation of the divergences
for $-1 \geq n>-3$ in the sum $N_0+M_0$ corresponds to the well-known cancellation between the full $P_{22}$ and $P_{13}$ contributions
in an EdS cosmology. As expected this cancellation generalizes to the gEdS case, but without the need for any cancellation in 
the $\alpha$-dependent terms. We will show below that the corrections in standard (e.g. LCDM-like) cosmologies to the one-loop 
PS are expressed in terms of the four infra-red safe integrals $M_1,M_2,N_1,N_2$.}
\label{Table1}
\end{table}

\label{section-PS-asymptotics}
We now consider the convergence properties of these one-loop contributions to the PS in the gEdS model i.e. for what asymptotic behaviours of 
the linear PS, $P_{\text{lin}}(k,a)$, the integrals converge. The results of this analysis are summarized in Table~\ref{Table1}. These are obtained 
straightforwardly by considering the $r \rightarrow 0$, and $r\rightarrow \infty$, behaviours of the integrands for the infra-red and ultra-violet limits, 
respectively. In the $M_i$ integrals there is also a divergence in the angular integral at $\mu=1$ for $r=1$,  corresponding to 
$|\bf{q}-\bf{k}|\rightarrow 0$, but because of the symmetry  of exchange of the variables $\bf{k}$ and $\bf{q}-\bf{k}$ 
in the integral (as noted e.g. in  \cite{makino1992}), its contribution is identical to that from $r \rightarrow 0$ and is thus taken into account 
by multiplying it by a factor of two.   

The leading contribution to $M_0$ as $r \rightarrow 0$ corresponds to $m_0=\mu^2$, which when integrated  over angle, and multiplied by two,
gives a factor of $4/3$ which can be seen to exactly cancel the corresponding leading contribution from $N_0$. These contributions to each integral 
are proportional to $\int P_{\rm lin} (k) dk$ i.e. to the variance of the
displacement field. As understood in early works on perturbation theory (e.g. \cite{Vishniac1983,scoccimarro1996loop}, see also \cite{peloso2013galilean} for a useful discussion),  
these are unphysical divergences if the system (as is the case here, just as in standard EdS) is Galilean invariant, and their cancellation is a consequence of this invariance. The next term in the expansion as $r\rightarrow 0$ does not cancel but is proportional to $\int P_{\rm lin} (k) k^2 dk$ i.e. to the variance of the fluctuation field (which is Galilean invariant). Their sum $M_0+N_0$ is then said to be ``infra-red safe'' in this case: the integral converges for any
$P_{\text{lin}}(k,a)$ which is integrable at $k \rightarrow 0$.

We see, on the other hand, that the integrals $M_1,M_2, N_1$ and $N_2$ are all infra-red safe: for $r \rightarrow 0$, the 
functions $m_1,m_2,n_1,n_2$ all have, after integration over angles, a leading behaviour $\sim r^2$. This means that infra-red safety of all
the $\alpha$-dependent contributions is obtained term by term, without any cancellation between different terms. This is evidently a necessary condition to obtain infra-red safety given that the $\alpha$-dependence in the $F_2$ and $F_3$ kernels are given through independent non-linear functions of $\alpha$.  
As we will see below it turns out that we can write the cosmological corrections (due to
departures from EdS) of the one-loop PS in LCDM-like models
in terms of the same  integrals. Infra-red safety of these cosmological corrections is thus explicit and 
their numerical calculation simplified, without the additional manipulations needed for the 
leading EdS approximated contribution (see e.g. \cite{carrasco20142}). 


Although we will not discuss them further in this paper, it is interesting to consider also the ultraviolet divergences. These are physical divergences marking a real breakdown of standard perturbation theory. It is straightforward to show (see Appendix \ref{Appendix-asymptotic behavior of the one-loop power spectrum} for details) that combining the leading contributions in the six integrals above
we obtain
\begin{equation}\label{eq:asymptotic_P22_big_q}
P_{22}(k,a)=\frac{7+36\alpha+92\alpha^{2}}{30(1+6\alpha)^{2}}k^{4}\int \frac{d^{3}q}{(2\pi)^3} \frac{P_{\text{lin}}^{2}(q)}{q^{4}} +\cdots,
\end{equation}
and 
 \begin{eqnarray}\label{eq:asymptotic_P13_big_q}
     2P_{13}(k)
    &=&\frac{7-14\alpha-176\alpha^{2}}{15(1+6\alpha)(1+8\alpha)} k^2 P_{\text{lin}}(k)\int \frac{d^{3}q}{(2\pi)^3} \frac{P_{\text{lin}}(q)}{q^{2}}\nonumber \\
    & &+\frac{12(8\alpha-1)(1+2\alpha)}{105(1+6\alpha)(1+8\alpha)} k^4 P_{\text{lin}}(k) \nonumber\\
    & &\times \int \frac{d^{3}q}{(2\pi)^3} \frac{P_{\text{lin}}(q)}{q^{4}} +\cdots
 \end{eqnarray}
where the dots now indicate terms proportional to integrals which converge more rapidly in the ultra-violet than these leading terms. The first expression corresponds to the sum of the leading contribution in the
integrals $M_i$, which we see (as indicated in Table~\ref{Table1}) diverge for $n \geq 1/2$, 
while the second shows the sum of the leading contributions from the $N_i$, the first diverging
for $n\geq -1$ (as indicated in Table~\ref{Table1}) and the next one diverging for $n\geq 1$.

The integrals in these expressions are the same as those in the standard EdS models, the only difference being in the coefficients which are manifestly all explicitly $\alpha$ dependent.
The leading divergence overall is that in $P_{13}$, diverging for $n \geq -1$.
We note that this remains true in the family of gEdS models, except that 
there is a specific value, $\alpha=0.1635\cdots$, at which the coefficient of this leading term, 
proportional to $7-14\alpha-176\alpha^{2}$, vanishes. For this specific value 
of $\alpha$ therefore the one-loop PS is ultra-violet convergent for 
$n < 1/2$. We will not explore further in this article the physical significance 
of this result, nor more generally the $\alpha$ dependence of the ultra-violet divergences.
To address these questions it is necessary to consider the regularisation of these divergences, using for example, the RPT or EFT approaches (see references cited in the introduction above). We will explore these issues in future work.

\section{LCDM approximated as an interpolation of gEdS}
\label{LCDM approximated as an interpolation of gEdS}

In the rest of this article we explore the relation of these models to standard (e.g. LCDM-like) cosmological models. The class of models considered are the FLRW models described at the beginning of Section \ref{Generalized EdS models}, i.e., with a clustering matter component and a smooth component with a generic equation of state.
Further to this we will assume only that cosmology is matter dominated at 
early times, i.e. that it is EdS, or gEdS if there remains a smooth matter-like component (e.g. massive neutrinos or matter-like dark energy).
Specifically we explore whether we can make use of the analytical results for perturbation theory in gEdS to calculate, exactly or approximately, results in standard models (see references further below). More precisely we wish to consider whether 
the functions, $Q(z)$ say, characterising the evolution of fluctuations in perturbation theory in an LCDM model as a function of redshift $z$ may be interpolated on the 
family of gEds models, so that we can write
\begin{equation}
\label{interpolation}
  Q(z) \approx Q_{gEdS} (\alpha= \alpha_{eff}(z)),
\end{equation}
where $\alpha_{eff}(z)$ is an effective growth exponent which can be calculated given the parameters of the LCDM model. 
Further we will determine whether the approximation 
\begin{equation}
\label{adiabatic-interpolation}
  Q (z) \approx Q_{gEdS} (\kappa^2= 1/\Omega_m(z)),
\end{equation}
is valid, i.e., whether we can take
\begin{equation}
\label{adiabatic-alpha}
\alpha_{eff}(z)=-\frac{1}{4}+\frac{1}{4}\sqrt{1+24 \Omega_m(z)},
\end{equation}
which means that we match both the instantaneous expansion rate and growth rate of the FLRW cosmology to that of a 
gEdS cosmology. We will say in this case that the interpolation is \textit{adiabatic}, because we expect such an 
approximation to be valid when we can neglect the temporal variation of the smooth component
relative to that of matter. Indeed we will see in all our equations below for effective growth exponents
that the approximation Eq.~(\ref{adiabatic-alpha}) becomes exact as $w_s \rightarrow 0$, which corresponds to this limit.

We will apply here this analysis up to third order in perturbation theory, which is what is needed to calculate the one-loop 
power spectrum. We have chosen for our analysis to follow closely that of  \cite{takahashi2008third}, who have calculated explicitly the 
corrections in LCDM and provided phenomenological fits to the relevant redshift dependent functions for standard type models. As we will show, our method of analysis in seeking to relate these results to those in gEdS, leads to a simplification of the formulation and calculations of the results of  \cite{takahashi2008third}, and other treatments
given in the literature (e.g. \cite{bernardeau1993skewness, fasiello2016nonlinear, garny2021loop, fasiello2022perturbation})
Further in so doing it provides greater insight into the reason why these cosmology-dependent corrections are typically so small, and how they depend on cosmological parameters.
We compare our numerical results in detail to the fits  of \cite{takahashi2008third} in Appendix ~\ref{Appendix-Takehashi}. 
For completeness we detail also a comparison 
of our analysis, and numerical results when possible,
with that of \cite{fasiello2022perturbation}
in Appendix~\ref{Appendix-comparison-to-Fasiello}, 
and with those of \cite{bernardeau1993skewness} and
\cite{garny2021loop} in Appendix~\ref{Appendix-comparison-to-Bernardeau}.
 

\subsection{Linear growth rate interpolated on gEdS}
By combining Eqs.~ \eqref{eq:density_fourier} and  \eqref{eq:euler_fourier} 
at linear order (i.e. neglecting the source terms $S_{\Tilde{\alpha}}$ and $S_{\Tilde{\beta}}$) we obtain easily the equation obeyed by 
the linear growth factor $D_1(a)$ which can be cast as 
\begin{eqnarray}
    \frac{d^2}{d^{2}\ln a^{2}}\frac{D_{1}}{a}&+&\Big(4+\frac{d\ln H}{d\ln a}\Big)\frac{d}{d\ln a}\frac{D_{1}}{a}\nonumber \\
   &+&\Big(3+\frac{d\ln H}{d\ln a}-\frac{3}{2}\Omega_{m}(a)\Big)\frac{D_{1}}{a}=0.\label{equation of D_1}
\end{eqnarray}
Writing 
\begin{equation}
\label{D1-integral}
    D_1= \exp \left(\int_{\ln a_0}^{\ln a}\alpha_{1} (a) \, d(\ln a_0)\right),
\end{equation}
where $a_0$ to be the reference time at which $D_1=1$, and 
\begin{equation}\label{def-alpha1-alpha2s}
    \alpha_{1}=\frac{d\ln D_{1}}{d\ln a},
\end{equation}
we have, trivially, an interpolation in the sense of Eq.~(\ref{interpolation}) taking $\alpha_{eff}(z)=\alpha_{1}$. 

\begin{figure*}
    \includegraphics[width=6cm,height=6cm]{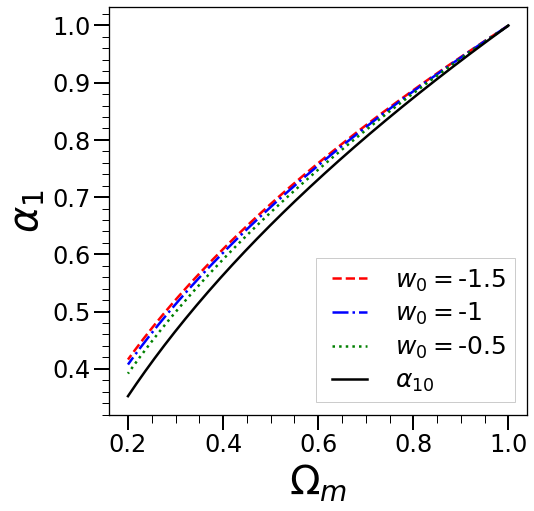}
    \includegraphics[width=6cm,height=6cm]{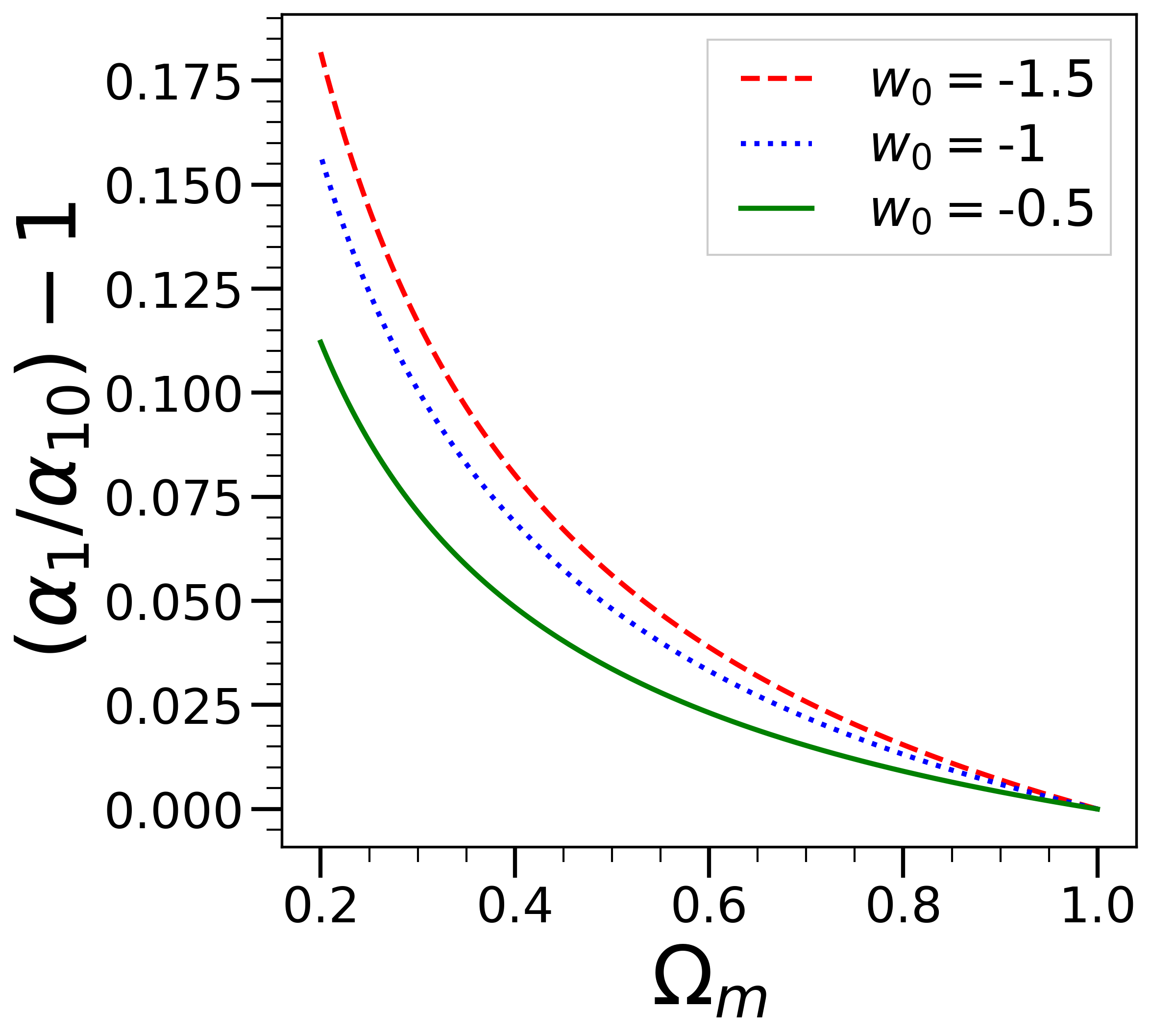}
    \caption{Left panel: Exact solution for the growth rate $\alpha_{1}$ compared with
its ``adiabatic gEdS  approximation'' $\alpha_{10}$, as a function of $\Omega_m$ and for different indicated values of $w_0$, parametrizing the constant equation of state of the smooth (dark energy) component. 
Right panel: the relative difference of the same quantities.} 
    \label{Figure-alpha1}
\end{figure*}

Let us now consider the accuracy of an adiabatic gEdS interpolation, in the sense
defined by Eq.~(\ref{adiabatic-interpolation}) above, i.e., the approximation
\begin{equation}
\label{alpha-adiabatic}
    \alpha_1(z)\approx \alpha_{10}(z)=\frac{1}{4} \big[-1+\sqrt{1+ 24 \Omega_{m}(z)}\big].
\end{equation}
Fig.~\ref{Figure-alpha1} shows a comparison of 
the exact growth rate $\alpha_1$ (obtained by solving Eq.~(\ref{equation of D_1}) 
numerically)  and $\alpha_{10}$, for models with constant equation of state $w_s=w_0$, for $w_0=-0.5,-1, -1.5$.
The left panel shows the two quantities, and the right panel their ratio.
We see that, even for the LCDM case ($w_0=-1$), the approximation is quite 
good down to $z=0$ (when $\Omega_m \approx 0.3$). 
This result is somewhat surprising as one would anticipate that it would 
hold only for $|w_0|$ small compared to unity. To understand what is observed better,
we re-express Eq.~(\ref{equation of D_1}) as
an equation for $\alpha_1$: 
\begin{equation}\label{alpha1}
    \left[\frac{d\alpha_{1}}{d\ln a}-\frac{3}{2}w \alpha_{1}\right]+\alpha_{1}^2
    +\frac{1}{2} \alpha_{1}
    -\frac{3}{2}\Omega_{m}=0,
\end{equation}
where 
\begin{equation}\label{eq-alpha1-alpha2AB}
 w=-1-\frac{2}{3}\frac{d\ln H}{d\ln a}=\frac{1}{3}\frac{d\ln\Omega_m}{d\ln a},
\end{equation}
and for the specific case of a smooth (dark energy) component with the constant equation of state, $w_s=w_0$, we have
\begin{equation}
    w=w_0(1-\Omega_m).
\end{equation}
The approximate solution  Eq.~(\ref{alpha-adiabatic}) corresponds to neglecting the first two terms in square brackets  i.e.,  setting $w=0$ and neglecting the time derivative, and taking the growing mode solution of the remaining quadratic equation. It is thus formally the solution at leading order in a gradient expansion in $\Omega_m(z)$ for which $w$ (or $w_0$ for a constant equation of state) is the control parameter. 
Given that e.g. in LCDM $w=-0.7$ when $\Omega_m=0.3$, it is indeed, as we have noted, surprising that the leading $w=0$ approximation is so good.
To see why this is the case more explicitly we
define 
\begin{equation}
\label{alpha1-interpolation-ansatz}
    \alpha_{1}=\alpha_{10}(1+\epsilon_{1} (z))\,.
\end{equation}
and rewrite Eq. (\ref{alpha1}), neglecting the single quadratic term in $\epsilon_1$, as a  
linearized equation in $\epsilon_1$: 
\begin{equation}\label{epsilon1_general}
    \frac{d\epsilon_{1}}{d\ln a}+\Big[-S (\Omega_{m}) w+(\frac{1}{2}+2\alpha_{10})\Big]\epsilon_{1}=S (\Omega_{m}) w,
\end{equation}
where 
\begin{equation}\label{S_omega_m}
    S (\Omega_{m})=\frac{3}{2}-\frac{1}{w}\frac{d\ln \alpha_{10}}{d\ln a}=\frac{3}{2}-\frac{9\Omega_{m}}{\alpha_{10}(1+4\alpha_{10})},
\end{equation}
is the coefficient multiplying $w$ which gives the source term for $\epsilon_1$, 
the correction to the leading adiabatic gEdS approximation.

\begin{figure}[b]
\includegraphics[width=7cm, height=7cm]{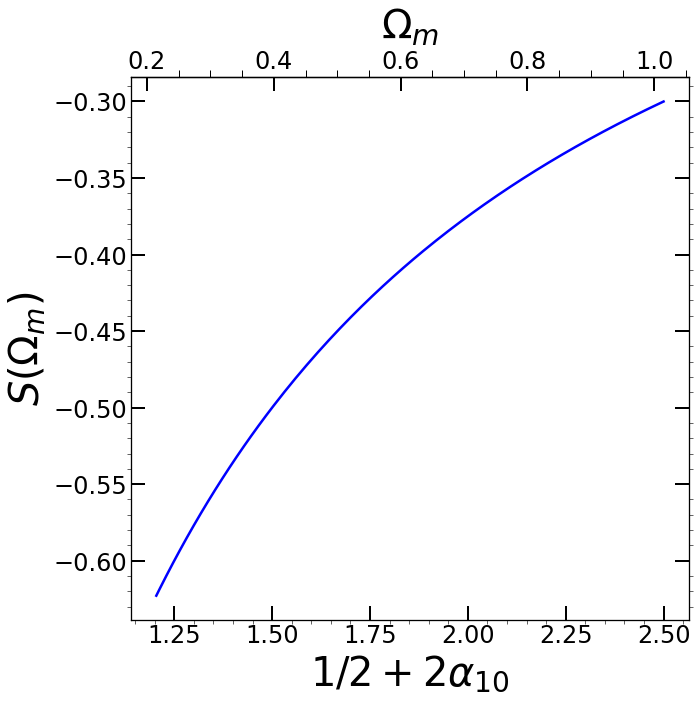}
\caption{The coefficient $S(\Omega_m)$ of the source term in Eq.~(\ref{epsilon1_general}) as a function of $(1/2+2\alpha_{10})$, and 
as a function of $\Omega_m$ (upper $x$-axis). We see that the ratio of $|S(\Omega_m)|/(1/2+2\alpha_{10})$ remains quite small as 
dark energy starts to dominate, so that even for $|w| \sim 1$ the adiabatic gEdS approximation for the linear growth factor 
can remain good.}
\label{S_Omega-vs-alpha10}
\end{figure}

Why $\epsilon_1$ remains small for relatively large $w$ is just a consequence of the smallness of the source term $S$ relative to the coefficient of the $\epsilon_1$ term on the left-hand size of the equation. This is shown explicitly in Fig. \ref{S_Omega-vs-alpha10} which shows the numerical value of $S(\Omega_m)$  plotted against that of $(\frac{1}{2}+2\alpha_{10})$. 
The relative smallness of $S$ arises from an approximate cancellation of the two terms in Eq.~(\ref{S_omega_m}), which 
corresponds to that between the two
terms in square brackets in Eq.~(\ref{alpha1}). Indeed
at high redshift we have 
$S \approx \frac{3}{2}-\frac{9}{5}=-0.3$.

\begin{figure}[b]
\includegraphics[width=7cm, height=7cm]{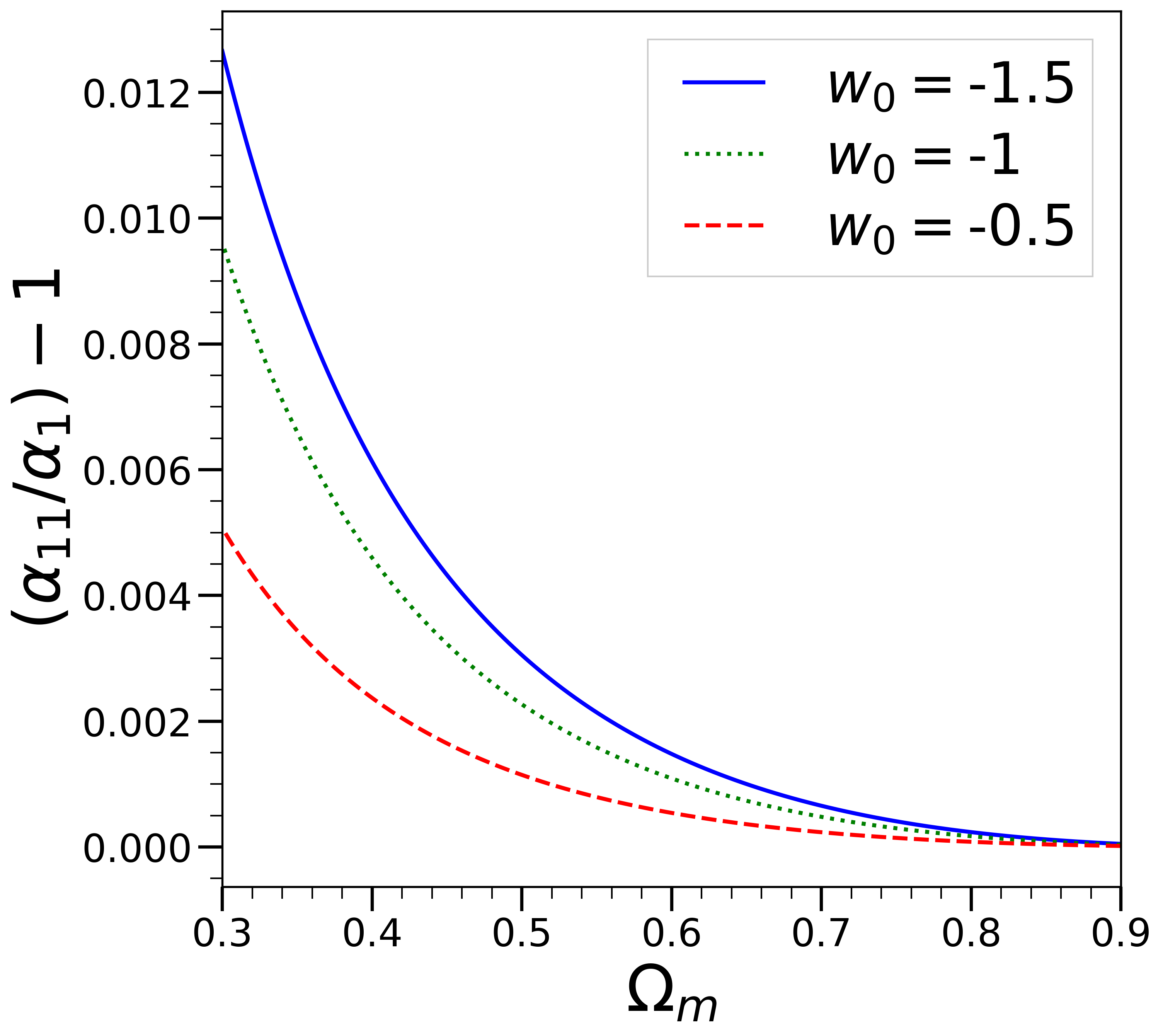}
\caption{Ratio of the growth rate $\alpha_{11}$ obtained 
using the linearized approximation Eq.~(\ref{epsilon1_general}) 
for $\epsilon_1=(\alpha_1/\alpha_{10})-1$, compared to the exact 
solution from Eq.~(\ref{alpha1}).}
\label{epsilon1-Omega}
\end{figure}

A corollary of this observed accuracy of the leading gEdS interpolation is that the approximation for $\alpha_1$ obtained
by solving the linear Eq.~(\ref{epsilon1_general})
is an accurate one. 
Fig. \ref{epsilon1-Omega} shows 
a comparison of the value of the growth 
rate obtained in this way with the exact
$\alpha_1$. We see that it is indeed
accurate well below the sub-percent level even for a LCDM model at $z=0$. 

\subsection{Second-order growth rates interpolated on gEdS models}

In treating LCDM models in perturbation theory at second and higher order, a considerable additional complexity arises because the separability of the EdS solution is not valid. In practice it has been shown (see e.g. \cite{Bernardeau2002, fasiello2016nonlinear, garny2021loop} 
and references therein) that the associated corrections are very small for standard type cosmological models, and for most purposes can be done using the EdS kernels and the assumption of separability, replacing simply the EdS linear theory growth rate by that of the model. Given the ever-improving precision of cosmological observations, however, even small theoretical errors arising from cosmology dependence are of interest and have been discussed in a number of recent works (see e.g. \cite{baldauf2013modelling,Baldauf2015bispectrum,aviles2020lagrangian, aviles2021clustering,garny2021loop,steele2021precise,baldauf2021two}).

We discuss here the calculations of such corrections in light of our results in gEdS models. More specifically we consider whether the time-dependent functions, additional to $D_1(a)$, which appear in describing the cosmology dependence of the evolved density field can be interpolated on gEdS models in the sense we have defined at the beginning of this section. In this subsection we 
consider the density fluctuation at second order, and in the following one at third order.
For conciseness, we will not present here the full derivation of the
results we use from EPT including cosmological corrections as it has been given in 
several other works, both for LCDM-type models (\cite{takahashi2008third,garny2021loop, fasiello2022perturbation})  or in a broader  class of models (e.g. \cite{fasiello2016nonlinear, taruya2016constructing}). As anticipated above we will follow in the
main text the treatment of \cite{takahashi2008third}, but we also provide 
in Appendices \ref{Appendix-comparison-to-Fasiello}-\ref{Appendix-comparison-to-Bernardeau} detailed comparison with several
other works
\cite{fasiello2022perturbation,bernardeau1993skewness,garny2021loop}.

The solution for the density perturbation in a general FLRW cosmology at second order is given by \cite{takahashi2008third} as
\begin{equation}\label{delta21}
    \delta_{2}(\textbf{k},a)=D_{2A} A (\textbf{k})+ D_{2B} B (\textbf{k}),
\end{equation}
where 
\begin{eqnarray}
    A (\textbf{k})=\frac{5}{7}\hat{A}(\textbf{k})=\frac{5}{7} \int d^{3}\textbf{q} \tilde{\alpha} (\textbf{q}, \textbf{k}-\textbf{q}) \delta_{1} (\textbf{q}) \delta_{1}(\textbf{k}-\textbf{q}),\nonumber\\
    B(\textbf{k})=\frac{2}{7}\hat{B}(\textbf{k})=\frac{2}{7}\int d^{3}\textbf{q} \tilde{\beta} (\textbf{q}, \textbf{k}-\textbf{q}) \delta_{1} (\textbf{q}) \delta_{1}(\textbf{k}-\textbf{q}), \nonumber\\
\end{eqnarray}
and $D_{2A}$ and $D_{2B}$ are time-dependent functions
which are solutions to the equations
\begin{eqnarray}\label{growth_D2}
            & &\frac{d^2}{d\ln a^{2}}\frac{D_2}{a^{2}}+\Big(6+\frac{d\ln H}{d\ln a}\Big)\frac{d}{d\ln a}\frac{D_{2}}{a^2}\nonumber\\
            & &+\Big[8+2\frac{d\ln H}{d\ln a}-\frac{3}{2}\Omega_{m}(a)\Big]\frac{D_{2}}{a^{2}}
      \nonumber \\&=&\begin{cases}
    \frac{7}{5}\Bigg[\Big(\frac{dD_1}{da}\Big)^2 +\frac{3}{2}\Omega_{m} (a)\Big(\frac{D_1}{a}\Big)^2\Bigg] \text{ for $D_{2A}$},\\
    \frac{7}{2} \Big(\frac{d D_1}{da}\Big)^2 \text{ for $D_{2B}$}.
  \end{cases}
\end{eqnarray}

These equations have been written in a form adapted to their comparison with the standard EdS model, for which the right-hand side vanishes giving the solutions $D_{2A}=D_{2B}=a^2$.
It is straightforward to verify that in the gEdS cosmology they 
admit the solutions
\begin{equation}
    D_{2A}=\frac{7}{5} \frac{1+4\alpha}{1+6\alpha} a^{2 \alpha} ,\quad D_{2B}=\frac{7}{2}\frac{2\alpha}{1+6\alpha} a^{2 \alpha},
    \label{gEds-D2solutions} 
\end{equation}
corresponding exactly to the separable solutions we have derived in the previous section.

In light of these solutions it is natural to define the functions
\begin{equation}
d_{2A}=\frac{5}{7}\frac{D_{2A}}{D_1^2}, \quad d_{2B}=\frac{2}{7}\frac{D_{2B}}{D_1^2},
\end{equation}
in terms of which
\begin{equation}\label{delta22}
    \delta_{2}(\textbf{k},a)=d_{2A} D_1^2 \hat{A} (\textbf{k})+d_{2B} D_1^2 \hat{B} (\textbf{k}) \,.
\end{equation}
When the functions $d_{2A}$
and $d_{2B}$ are constant in time,  $\delta_{2}(\textbf{k},a)$ is a time-independent functional of 
$\delta_{1}(\textbf{k},a)$.
Non-separability, on the contrary, corresponds to a time-dependence of the functions $d_{2A}$
and $d_{2B}$. 
We first note that Eqs.~(\ref{growth_D2}) can be written, making use of the definitions of $\alpha_1$ and $w$ in the previous section, and using Eq.~(\ref{alpha1}) to eliminate terms in 
$d\ln \alpha_1/d\ln a$, as 
\begin{equation}\label{growth_d2}
{\cal {D}}^{(2)} d_{2A}=\alpha_{1}^2 +\frac{3}{2}\Omega_{m},\quad {\cal {D}}^{(2)} d_{2B}=\alpha_{1}^2  ,
\end{equation}
where 
\begin{equation}\label{D2-operator}
        {\cal {D}}^{(2)}=\frac{d^2}{d\ln a^{2}}+\Big[\frac{1}{2}(1-3w)+4\alpha_{1}\Big]\frac{d}{d\ln a}+
        (2\alpha_{1}^2+\frac{3}{2}\Omega_{m}).
\end{equation}
We see immediately that ${\cal {D}}^{(2)} (d_{2A}+d_{2B}-1)=0$, from which it follows 
that the evident property 
\begin{equation}\label{relation-d2AB}
d_{2A}+d_{2B}=1,
\end{equation}
of the usual EdS solution  (with $d_{2A}=5/7$ and $d_{2B}=2/7$), shared by the gEdS cosmologies (with $d_{2A}=\frac{1+4\alpha}{1+6\alpha}$ and $d_{2B}=\frac{2\alpha}{1+6\alpha}$), will hold in a general LCDM-like cosmology, if at asymptotically early times the cosmology approaches EdS (or gEdS).    
There is therefore in this case only one real function (in addition to $D_1$) is needed to describe the evolution of the density perturbation at second order, just as in the original treatment of
\cite{bernardeau1993skewness}. Given our gEdS solutions  (\ref{gEds-D2solutions}), it is natural to define, {\it without loss of generality}, the time dependence using the function $\alpha_{2}(a)$ 
defined by
\begin{equation}
    d_{2A}(a)=\frac{1+4\alpha_{2}(a)}{1+6\alpha_{2}(a)}, \quad d_{2B}(a)=1-d_{2A}(a)\,.
    \label{def-alpha-eff-2} 
\end{equation}
$\alpha_{2}(a)$ is just then the value of $\alpha$ in the gEdS model with the same instantaneous values of  $d_{2A}$ and $d_{2B}$, i.e., 
it is the effective growth exponent required to interpolate (exactly) these functions on gEdS models. Anticipating the fact that we will consider in  particular here the small corrections due to the non-separability of the LCDM model relative to the EdS approximation (with $d_{2A}=5/7$ and $d_{2B}=2/7$), we also define 
\begin{equation}
    d_{2A}
    =\frac{5}{7}\Big[1+\frac{2}{35}\gamma_2\Big], \quad
    d_{2B}
    =\frac{2}{7}\Big[1-\frac{1}{7}\gamma_2\Big],
    \label{def-gamma} 
\end{equation}
where 
\begin{equation}
    \gamma_2=\frac{1-\alpha_{\rm 2}}{1+\frac{6}{7}(1-\alpha_{\rm 2})}\,,
    \label{alphaeff-gamma} 
\end{equation}
i.e. the coefficients in the definition of the perturbation $\gamma_2$ have been chosen so that, in the limit of small $\gamma_2$, we obtain $\gamma_2=1-\alpha_{2}$. 

From Eq.~(\ref{growth_d2}) we have that the equation obeyed by $\gamma_2$ is 
\begin{equation}\label{eq_gamma2}
{\cal {D}}^{(2)} \gamma_{2}=\frac{21}{2} \Big(\Omega_{m}-\alpha_{1}^2\Big).
\end{equation}
It is instructive to rewrite this equation
as 
\begin{eqnarray} 
\label{evolution_gamma2}
    \frac{d^2 \gamma_{2}}{d\ln a^2} &+&\Big[\frac{1}{2}(1-3w) + 4\alpha_{1}\Big]\frac{d \gamma_{2}}{d\ln a} \nonumber 
    \\&+&\Big(2\alpha_{1}^2+\frac{3}{2}\Omega_{m}\Big) [\gamma_2- \gamma_2^{(0)}]
    =0,
\end{eqnarray} 
where 
\begin{equation} 
\label{gamma2_0}
  \gamma_2^{(0)} 
    =\frac{21(\Omega_{m}-\alpha_{1}^2)}{4\alpha_{1}^2+3\Omega_{m}}\,.
\end{equation}
We recover explicitly from this equation the result  \cite{martel371second,scoccimarro1998nonlinear} that the solution is separable if and only if the ratio $\Omega_{m}/\alpha_{1}^2$ is constant. Indeed, as we have noted, separability corresponds to a time-independent solution for $\gamma_2$, which here is admitted if and only if $\gamma_2^{(0)}$ is constant in time, and therefore if and only if
the ratio $\Omega_{m}/\alpha_{1}^2$ is so. More generally we note that solving 
in the adiabatic gEdS approximation, i.e. neglecting all terms proportional to $w$ 
and all time derivatives, and taking, in the same approximation, $\alpha_1=\alpha_{10}$, we then obtain at leading order
\begin{equation}
\label{gamma2-adiabatic}
\gamma_2=7\frac{1-\alpha_{10}}{1+6 \alpha_{10}},
\end{equation}
which corresponds to 
\begin{equation}
\label{alpha2-adiabatic}
\alpha_{\rm 2}=\alpha_{10},
\end{equation}
i.e. we recover as expected, as $w\rightarrow 0$, the adiabatic gEdS approximation for $d_{2A}$ (and $d_{2B}$).

\begin{figure}[b]
\includegraphics[width=7.5cm, height=7.5cm]{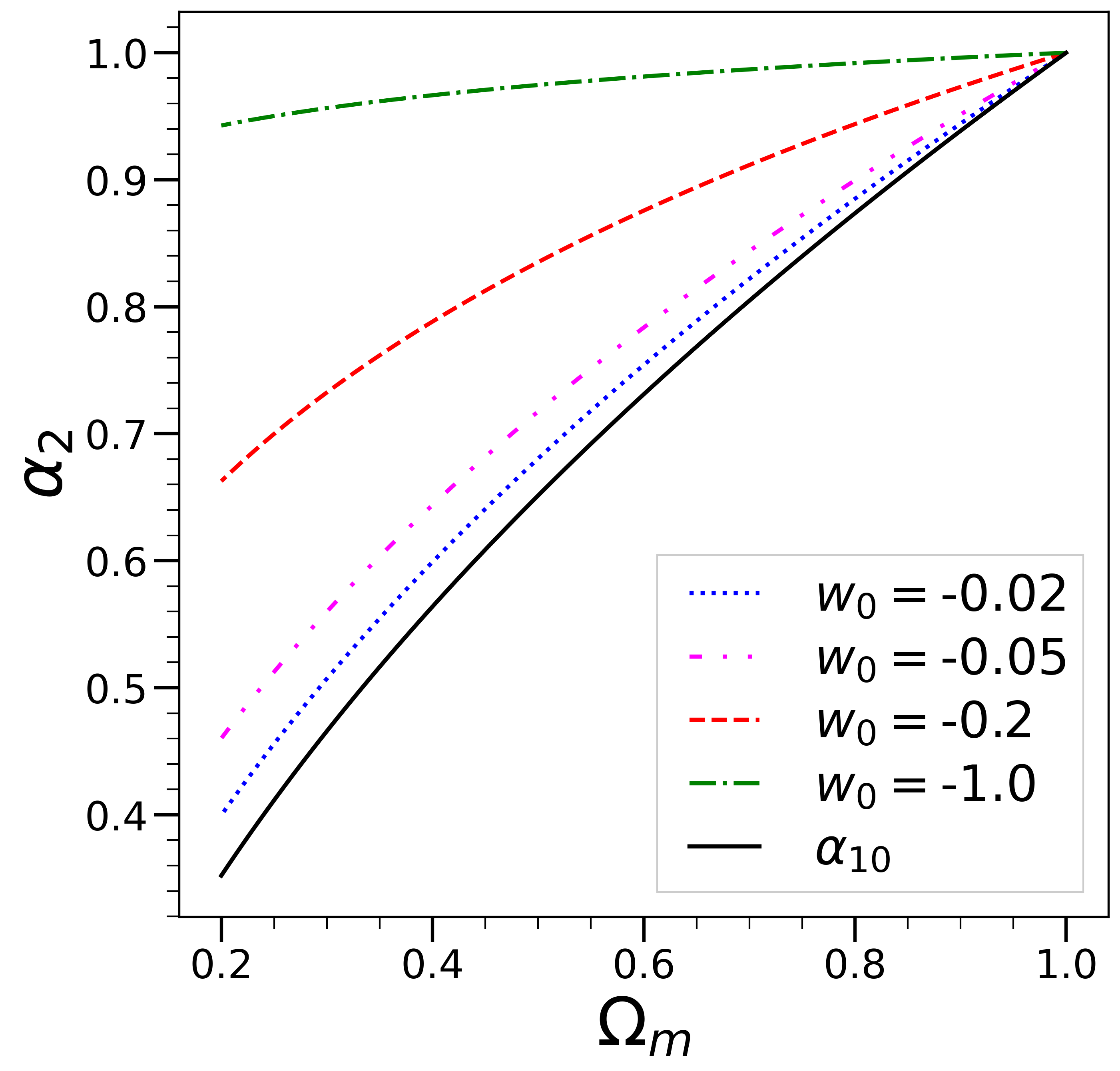}
\caption{The effective growth rate $\alpha_2$ 
of a gEdS model with the same instantaneous value of the functions $d_{2A}$ (and $d_{2B}$), for dark energy models with the constant equation of state parametrized by the indicated values of $w_0$. We see that, for very small absolute values of 
$w_0$, $\alpha_2$ is well approximated by the adiabatic linear growth rate
$\alpha_{10}$ so that, as expected, $d_{2A}$ and $d_{2B}$ are well approximated by an adiabatic interpolation of gEdS models. For larger absolute values of $w_0$, $\alpha_2$ stays much closer to its initial value because of rapid evolution which causes effective damping in its evolution equation.}
\label{alpha2_and_alpha10}
\end{figure}

Fig.~\ref{alpha2_and_alpha10} shows the numerical solutions obtained for $\alpha_2$
as a function of $\Omega_m$, for models with the different indicated values of $w_0$. 
These are obtained by solving Eq.~(\ref{eq_gamma2}) for $\gamma_2$, with initial conditions corresponding to the asymptote to EdS as $z \rightarrow \infty $: 
\begin{equation}
\gamma_2(0)=0,\quad    
\frac{d \gamma_2}{d \ln a}(0)=0\,. 
\end{equation}
In practice, rather than extrapolating to an initial time $a_i \ll 1$ at which  $\Omega_m\approx 1$, we can avoid the very early time transients by taking directly as the initial condition 
\begin{equation}
\gamma_2(a_i)=1-\alpha_{10}(a_i),\quad    
\frac{d \gamma_2}{d \ln a}(a_i)=-\frac{d\alpha_{10}}{d \ln a}(a_i)\,,
\end{equation}
since $\gamma_2(a)=1-\alpha_{10}(a)$, as we have noted above, becomes
a good approximation to the exact evolution in the limit $w \ll 1$. 
Indeed this can be seen in Fig.~\ref{alpha2_and_alpha10}, in which  
this adiabatic approximation $\alpha_2=\alpha_{10}$ is shown (solid line). 
We see that,  for very small absolute values of $w_0$, $\alpha_2$ is very well approximated by $\alpha_{10}$. In other words as expected, $d_{2A}$ and $d_{2B}$ are then well approximated by an adiabatic interpolation of gEdS models. 
However, unlike what we observed for $\alpha_1(z)$, this adiabatic approximation fails for increasing values of $|w_0|$. In this case we see that $\alpha_2$ stays much closer to its initial value. This can be understood 
simply as a result of the increasing contribution of the dark energy, which increases (through the dependence in $w$) an effective damping term in Eq.~(\ref{eq_gamma2})  which slows greatly the ``relaxation'' to $\gamma_2=\gamma_2^{(0)}$ occurring for small $|w_0|$.


\subsection{Third order growth rates}
Following again the treatment of \cite{takahashi2008third}, we write the density perturbation to third-order as 
\begin{eqnarray} \label{third-order-density-field}
        \delta_{3} (\textbf{k}, a) &=&D_{3AA} (a) C_{AA} (\textbf{k}) + D'_{3AA} (a) C'_{AA} (\textbf{k}) 
        \nonumber \\
        & &+ D_{3AB} (a) C_{AB} (\textbf{k})+ D'_{3AB} (a) C'_{AB} (\textbf{k}) \nonumber \\
        & &+ D_{3BA} (a) C_{BA} (\textbf{k}) + D_{3BB} (a) C_{BB} (\textbf{k}),\nonumber \\
\end{eqnarray} 
where 
\begin{eqnarray}
C_{AA} (\textbf{k}) &=&\frac{7}{18}\int d^{3}\textbf{q} \Tilde{\alpha} (\textbf{q}, \textbf{k}-\textbf{q}) \delta_{1} (\textbf{q}) A(\textbf{k}-\textbf{q}),\nonumber\\
\\
C_{AA}^\prime (\textbf{k}) &=&\frac{7}{30}\int d^{3}\textbf{q} \Tilde{\alpha} (\textbf{q}, \textbf{k}-\textbf{q}) \delta_{1} (\textbf{k}-\textbf{q}) A(\textbf{q}),\nonumber\\\\
C_{AB}(\textbf{k}) &=&\frac{7}{18}\int d^{3}\textbf{q} \Tilde{\alpha} (\textbf{q}, \textbf{k}-\textbf{q}) \delta_{1} (\textbf{q}) B(\textbf{k}-\textbf{q}),\nonumber\\\\
C_{AB}^\prime(\textbf{k}) &=&\frac{7}{9}\int d^{3}\textbf{q} \Tilde{\alpha} (\textbf{q}, \textbf{k}-\textbf{q}) \delta_{1} (\textbf{k}-\textbf{q}) B(\textbf{q}),\nonumber\\\\
C_{BA}(\textbf{k}) &=&\frac{2}{15}\int d^{3}\textbf{q} \Tilde{\beta} (\textbf{q}, \textbf{k}-\textbf{q}) \delta_{1} (\textbf{q}) A(\textbf{k}-\textbf{q}),\nonumber\\\\
C_{BB} (\textbf{k}) &=&\frac{4}{9}\int d^{3}\textbf{q}\Tilde{\beta} (\textbf{q}, \textbf{k}-\textbf{q}) \delta_{1} (\textbf{q}) B(\textbf{k}-\textbf{q}).\nonumber\\
\end{eqnarray}
and the six functions $D_{3XX}$ are the
solutions of the four differential equations 
\begin{eqnarray}\label{growth_D}
            & &\frac{d^2}{d\ln a^{2}}\frac{D_3}{a^{3}}+\Big(8+\frac{d\ln H}{d\ln a}\Big)\frac{d}{d\ln a}\frac{D_{3}}{a^3}
            \nonumber\\
            & &+\Big[15+3\frac{d\ln H}{d\ln a}-\frac{3}{2}\Omega_{m}(a)\Big]\frac{D_{3}}{a^{3}}\nonumber
       \\&=&\begin{cases}
    \frac{18}{7}\Big[2\frac{dD_1}{da} +\frac{3}{2}\Omega_{m} (a)\frac{D_1}{a}\Big]\frac{D_{2A,B}}{a^2}\\+\frac{18}{7}a\frac{d D_1}{d a}\frac{d}{da}\frac{D_{2A,B}}{a^2}\text{ for $D_{3AA,3AB}$}\\
    15 \frac{d D_1}{d a} \Big[a\frac{d}{d a}\frac{D_{2A}}{a^2}+2\frac{D_{2A}}{a^2}-\frac{7}{5}\frac{D_1}{a}\frac{d D_1}{d a}\Big] \text{ for $D_{3BA}$}\\
    \frac{9}{2} \frac{d D_1}{d a} \Big[a\frac{d}{d a}\frac{D_{2B}}{a^2}+2\frac{D_{2B}}{a^2}\Big] \text{ for $D_{3BB}$},
  \end{cases}
\end{eqnarray}
and the two remaining functions are determined by the relations 
\begin{eqnarray}
\frac{5}{18}D_{3AA}+ \frac{2}{9}D_{3AB}^\prime=\frac{1}{2}D_{1}^{3}, \nonumber \\ 
\label{constraints}
\frac{1}{6}D_{3AA}^\prime+ \frac{1}{9}D_{3AB}+\frac{2}{21}D_{3BA}+\frac{8}{63}D_{3BB}=\frac{1}{2}D_{1}^{3}.\nonumber\\
\end{eqnarray}

Analogously to the previous section we now define
\begin{eqnarray}\label{relation-d3xx}
d_{3AA}&=& \frac{5}{9} \frac{D_{3AA}}{D_1^3}, \quad d_{3AA}^\prime= \frac{1}{3} \frac{D_{3AA}^\prime}{D_1^3},\nonumber\\
d_{3AB}&=& \frac{2}{9} \frac{D_{3AB}}{D_1^3}, \quad d_{3AB}^\prime= \frac{4}{9} \frac{D_{3AB}^\prime}{D_1^3}, \nonumber\\
\quad d_{3BA}&=& \frac{2}{21} \frac{D_{3BA}}{D_1^3},\quad d_{3BB}= \frac{8}{63} \frac{D_{3BB}}{D_1^3}.    
\end{eqnarray}

For the case of the gEdS cosmologies, it is simple to verify using the kernels $F_2$, $G_2$ and $F_3$ derived in Section~\ref{Perturbation Theory kernels in generalized EdS models} that we obtain 
\begin{eqnarray}
\label{d3XX-gEdS}
d_{3AA} &=&d_3 d_2=\frac{1+4\alpha}{1+8\alpha}, \nonumber\\ 
d^\prime_{3AA} &=& d_3 \tilde{d}_2=\frac{1+2\alpha}{1+8\alpha},\nonumber\\ 
d_{3AB} &=&d_3 (1-d_2)=\frac{2\alpha}{1+8\alpha}, \nonumber\\ 
d^\prime_{3AB} &=& d_3(1-\tilde{d}_2)=\frac{4\alpha}{1+8\alpha},\nonumber \\ 
d_{3BA} &=&(1-d_3)\tilde{d}_2=\frac{2\alpha(1+2\alpha)}{(1+6\alpha)(1+8\alpha)}, \nonumber \\ 
d_{3BB} &=&(1-d_3)(1-\tilde{d}_2)=\frac{8\alpha^2}{(1+6\alpha)(1+8\alpha)}.\nonumber\\
\end{eqnarray}

It is straightforward to show that using
Eqs.~(\ref{growth_D}), and $d_{2B}=1-d_{2A}$,   
we obtain 
\begin{eqnarray}\label{growth_D3}
        {\cal {D}}^{(3)} d_{3AA}&=&
       (4\alpha_{1}^2 +3\Omega_{m}) d_{2A} + 2\alpha_1 \frac{d\, d_{2A}}{d\ln a},\nonumber\\
       {\cal {D}}^{(3)} d_{3AB}&=&(4\alpha_{1}^2 +3\Omega_{m})-(4\alpha_{1}^2 +3\Omega_{m}) d_{2A} \nonumber\\
       & &- 2\alpha_1 \frac{d\, d_{2A}}{d\ln a},\nonumber\\
       {\cal {D}}^{(3)} d_{3BA}&=&-2\alpha_1^2+4\alpha_{1}^2 d_{2A} + 2\alpha_1 \frac{d\, d_{2A}}{d\ln a},\nonumber\\
       {\cal {D}}^{(3)} d_{3BB}&=&4\alpha_{1}^2-4\alpha_{1}^2 d_{2A} - 2\alpha_1 \frac{d\, d_{2A}}{d\ln a},
\end{eqnarray}
where 
\begin{equation}\label{D3-operator1}
        {\cal {D}}^{(3)}=\frac{d^2}{d\ln a^{2}}+\Big[\frac{1}{2}(1-3w)+6\alpha_{1}\Big]\frac{d}{d\ln a}+
        3(2\alpha_{1}^2+\Omega_{m})\,.
\end{equation}
The two constraints Eqs.~(\ref{constraints}) may be written as 
\begin{eqnarray}\label{constraint_d3xx}
d_{3AA}+ d_{3AB}^\prime=1, \\ 
d_{3AB}+ d_{3AA}^\prime +2(d_{3BA}+ d_{3BB})=1.
\label{constraint_d3xx_2}
\end{eqnarray}
Setting all derivatives to zero in
Eq.~(\ref{growth_D3}), we recover, as required, the expression 
for the gEdS cosmologies Eq.~(\ref{d3XX-gEdS})
above.

Taking the sum of Eqs.~(\ref{growth_D3}) we note that we obtain also
\begin{equation}\label{constraint-eq-sum41}
        {\cal {D}}^{(3)} (d_{3AA}+ d_{3AB} +d_{3BA}+ d_{3BB}-1)=0,
\end{equation}
and therefore, given that this quantity is zero in an EdS (and gEdS) cosmology, 
we obtain the additional constraint 
\begin{eqnarray}
\label{constraint-one}
d_{3AA}+ d_{3AB} +d_{3BA}+ d_{3BB}=1,
\end{eqnarray}
when we assume that the cosmology is asymptotically EdS (or gEdS). Further writing Eq.~(\ref{growth_d2}) for $d_{2A}$ as 
\begin{equation}\label{growth_d2A}
{\cal {D}}^{(3)} d_{2A}=\alpha_{1}^2 +\frac{3}{2}\Omega_{m}+(4\alpha_{1}^2 +\frac{3}{2}\Omega_{m}) d_{2A} + 2\alpha_1 \frac{d\, d_{2A}}{d\ln a},
\end{equation}
we see also that
\begin{equation}\label{constraint-eq-sum4}
        {\cal {D}}^{(3)} (2d_{2A}-d_{3AA}+ d_{3BB}-1)=0,
\end{equation}
so that we can infer (again assuming the cosmology to be asymptotically EdS or gEdS)
the additional relation
\begin{equation}\label{constraint-eq-sum-4}
       2d_{2A}-d_{3AA}+ d_{3BB}=1.
\end{equation}
The full set of constraints now reads
\begin{eqnarray}
\label{constraints-all1}
d_{3AA}+ d_{3AB}^\prime=1 , \\ 
\label{constraints-all2}
d_{3AB}^\prime+ d_{3AA}^\prime +d_{3BA}+ d_{3BB}=1,\\
\label{constraints-all3}
d_{3AB}+ d_{3AA}+d_{3BA}+ d_{3BB}=1,\\
\label{constraints-all4}
2d_{2A}-d_{3AA}+d_{3BB}=1.
\end{eqnarray}
We can thus conclude that the perturbation at the third order can therefore 
be described fully by only just {\it two} independent functions in addition to $d_{2A}$. 
This result agrees again, as we saw at second order, with the original analysis of \cite{bernardeau1993skewness} for the asymptotic EdS case.
The exact mapping between our functions and 
those of \cite{bernardeau1993skewness} is given in 
Appendix~\ref{Appendix-comparison-to-Bernardeau}. 
Our analysis above shows that this same result
(and either our parametrization or that of 
\cite{bernardeau1993skewness}) generalizes to the 
broader class of cosmologies which are
asymptotically gEdS.

We choose now (arbitrarily) to use, in addition to $d_{2A}$, the two functions $d_{3AA}$ and $d_{3AB}$ to describe the
cosmology-dependent corrections to the 
third-order perturbation
To describe them 
in terms of interpolation on gEdS models we 
thus introduce, following the treatment of the second order perturbation, effective growth exponents as follows: 
  \begin{eqnarray}
    d_{3AA}(a)&=&\frac{1+4\alpha_{3AA}(a)}{1+8\alpha_{3AA}(a)}=\frac{5}{9}\Big[1+\frac{4}{45}\gamma_{3AA}(a)\Big], \nonumber\\  
    d_{3AB}(a)&=&\frac{2\alpha_{3AB}(a)}{1+8\alpha_{3AB}(a)}=\frac{2}{9}\Big[1-\frac{1}{9}\gamma_{3AB}(a)\Big], \nonumber \\ 
    \label{def-gammas-3} 
\end{eqnarray}
where, as for the second order case, the numerical coefficients have been chosen so that we will have, in the limit of small $\gamma_{3XX}$, that
$\alpha_{3XX} \approx 1-\gamma_{3XX}$. 

The evolution of the $\gamma_{3AA}$ and $\gamma_{3AB}$ are now
given by the equations
\begin{eqnarray} 
\label{evolution_gamma3XX_v0}
    {\cal {D}}^{(3)} \gamma_{3AA}&=&
    \frac{135}{14} \Big(\Omega_{m}-\alpha_{1}^2\Big)
       \nonumber\\ 
       & & +\frac{81}{98}\Big[(4\alpha_{1}^2 +3\Omega_{m}) \gamma_{2} + 2\alpha_1 \frac{d\, \gamma_{2}}{d\ln a}\Big],\ \\
          {\cal {D}}^{(3)} \gamma_{3AB}&=&
    -\frac{54}{7} \Big(\Omega_{m}-\alpha_{1}^2\Big)
       \nonumber\\ 
       & &+\frac{81}{49}\Big[(4\alpha_{1}^2 +3\Omega_{m})\gamma_{2} + 2\alpha_1 \frac{d\, \gamma_{2}}{d\ln a}\Big],\ \nonumber  \\
\end{eqnarray}
which can also be written as 
\begin{equation} 
\label{evolution_gamma3XX_v1}
   \begin{split}
    \frac{d^2 \gamma_{3XX}}{d\ln a^2} +&\Big[\frac{1}{2}(1-3w) +6\alpha_{1}\Big]\frac{d \gamma_{3XX}}{d\ln a} +3\Big[2\alpha_{1}^2+\Omega_{m}\Big] \\
    &\times (\gamma_{3XX}- \gamma_{3XX}^{(0)})=c_{3XX} \Big[(4\alpha_{1}^2 +3\Omega_{m}) \\
    &\times (\gamma_{2}-\gamma_2^{(0)}) + 2\alpha_1 \frac{d\, \gamma_{2}}{d\ln a}\Big],
   \end{split}
\end{equation}
where $c_{3AA}=\frac{81}{98}, c_{3AB}=\frac{81}{49}$ and 
\begin{equation}
\gamma_{3AA}^{(0)}=\gamma_{3AB}^{(0)}=\frac{9(\Omega_{m}-\alpha_{1}^2)}{2\alpha_{1}^2+\Omega_{m}}.
\end{equation}
For purposes of comparison it is useful to rewrite Eq.~(\ref{evolution_gamma2}) 
for $\gamma_2$ as 
\begin{equation} 
\label{evolution_gamma2_final_v0}
   \begin{split}
    {\cal {D}}^{(3)} \gamma_{2}&=
    \frac{21}{2} \Big(\Omega_{m}-\alpha_{1}^2\Big)
       +\Big[(4\alpha_{1}^2 +\frac{3}{2}\Omega_{m}) \gamma_{2} + 2\alpha_1 \frac{d\, \gamma_{2}}{d\ln a}\Big],
   \end{split}
\end{equation}
and 
\begin{equation} 
\label{evolution_gamma2_final-v1}
   \begin{split}
    \frac{d^2 \gamma_{2}}{d\ln a^2} +&\Big[\frac{1}{2}(1-3w) + 6\alpha_{1}\Big]\frac{d \gamma_{2}}{d\ln a} +3\Big[2\alpha_{1}^2+\Omega_{m}\Big]\\ &\times (\gamma_{2}-\gamma_{2}^{(0)})=\Big[(4\alpha_{1}^2 +\frac{3}{2}\Omega_{m}) (\gamma_{2}-\gamma_2^{(0)}) \\
    &+ 2\alpha_1 \frac{d\, \gamma_{2}}{d\ln a}\Big].
   \end{split}
\end{equation}
Solving Eqs.~(\ref{evolution_gamma2_final_v0}) and (\ref{evolution_gamma3XX_v0})
in the adiabatic  gEdS approximation, i.e. neglecting all time derivatives in these equations, we can see that we obtain the solution Eq.~(\ref{gamma2-adiabatic}) 
for $\gamma_2$ and now also 
\begin{equation}
\label{gamma2-adiabatic11}
\gamma_{3AA}=\gamma_{3AB}=9\frac{1-\alpha_{10}}{1+8\alpha_{1,0}} ,
\end{equation}
which corresponds, as anticipated, to the leading adiabatic solutions
\begin{equation}
\label{adiabatic-3alpha}
\alpha_{2} (a)=\alpha_{3AA}(a)=\alpha_{3AB}(a)=\alpha_{10}(a).
\end{equation}

\begin{figure}[t]
\includegraphics[width=7.5cm, height=7.5cm]{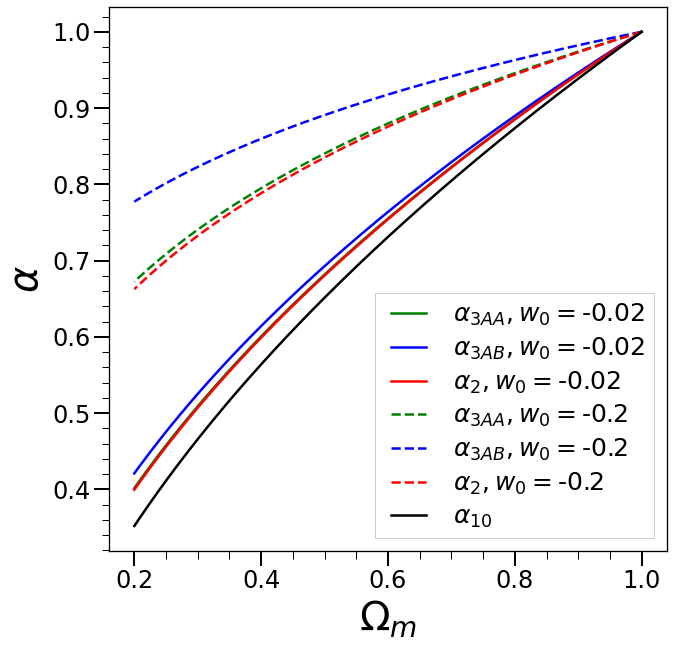}
\caption{Evolution of the effective growth rate $\alpha_2$, $\alpha_{3AA}$ and $\alpha_{3AB}$ 
which describes in general the cosmology-dependent corrections to the separable EdS approximation at third order in perturbation theory. The different sets of curves correspond to the different indicated values of $w_0$. Also shown is the adiabatic linear growth rate $\alpha_{10}$. We see that for
the smallest values the adiabatic approximation is good, but that it degrades increasingly as $w$ increases.}
\label{alpha_10_alpha_2_alpha_3_small_w}
\end{figure}

\begin{figure}[t]
\includegraphics[width=7cm, height=7cm]{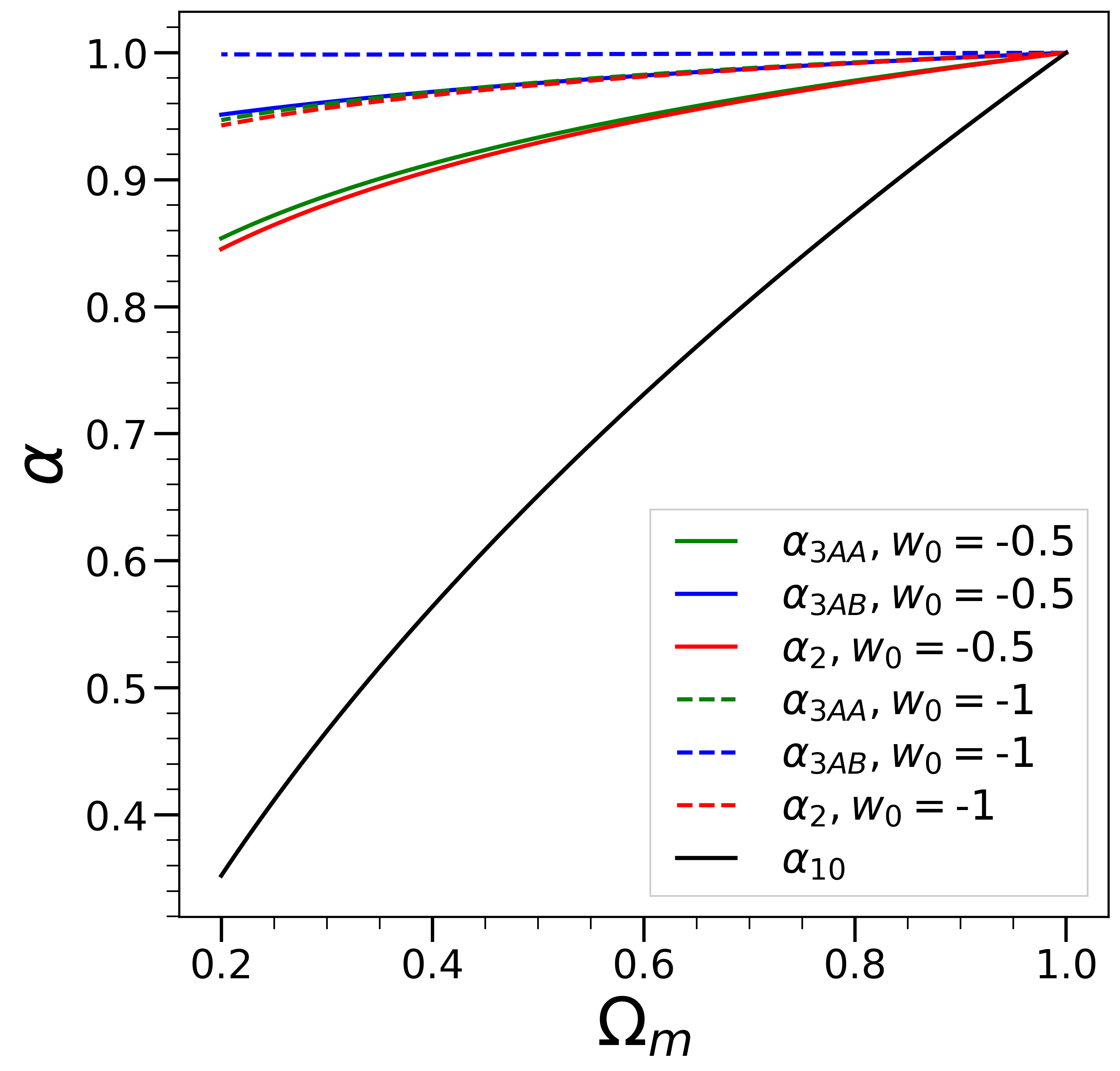}
\caption{Same as in the previous figure but for models with larger absolute values of 
$w_0$. We see that for these values of $w_0$ becomes larger, one of the three exponents has quite a different value from the $\alpha_2$, so an interpolation of
the functions describing deviations from separability on a single gEdS
the model with $\alpha=\alpha_2$ is not necessarily a valid approximation in general,
although it turns out to be for the calculation of the one-loop PS.}
\label{alpha_10_alpha_2_alpha_3_big_w}
\end{figure}

Fig.~\ref{alpha_10_alpha_2_alpha_3_small_w} and 
Fig.~\ref{alpha_10_alpha_2_alpha_3_big_w} show the results for the three effective growth exponents $\alpha_2$ (as in  Fig.~\ref{alpha2_and_alpha10}), $\alpha_{3AA}$
and $\alpha_{3AB}$ as a function of $\Omega_m$ obtained by numerically 
integrating these equations for different dark energy models with the fixed equation of state with the same initial conditions as for $\gamma_2$ (corresponding to EdS at
high redshift). For the smallest value of $w_0$ we see, as expected, all three exponents well approximated by the adiabatic solution Eq.~(\ref{adiabatic-3alpha}).
Then as $|w_0|$ increases we see that not only do the exponents differ from this solution but they start differing, albeit to a lesser extent, from one another.
More specifically we note that $\alpha_{2} \approx \alpha_{3AA}$, while 
the third exponent differs. The origin of these behaviours can be seen easily in the coefficients of the equations above: those in the  
equations for $\gamma_2$ and  $\alpha_{3AA}$ are almost identical, while 
those in the equation  $\alpha_{3AB}$ differ not only in magnitude (by about a factor of $2$) but also in sign.
In Appendix \ref{Appendix-Takehashi} we report a detailed comparison 
of our numerical results with those of \cite{takahashi2008third}, and adapt the fitting functions 
provided by it to infer corresponding fits for the two functions $\gamma_{3AA}$ and $\gamma_{3AB}$.

\subsection{The PS at one loop}\label{PS-one-loop}

We now derive our simplified expressions for the PS at one loop in the class of FLRW cosmologies specified in Section \ref{Generalized EdS models} (clustering matter plus a generic dark energy fluid component), making use of the
reduction of the number of independent functions and expressing these in terms of the 
effective growth exponents we have defined above. We also consider finally 
to what accuracy this exact result can be approximated by a direct interpolation 
of the PS in gEdS models (i.e. by just replacing $\alpha$ in the analytic gEdS 
kernels by a single effective growth rate). In our analysis we start from the
expressions we need (for each of the `22' and `13' contributions) in a form
easily comparable (up to notation) to the expressions given in \cite{takahashi2008third}.
This facilitates the explanation of the simplifications we obtain. 

\subsubsection{`22' term}
Starting from Eq.~(\ref{delta21}), 
we obtain the full time-dependent $P_{22}$ directly as
\begin{eqnarray}
P_{22}&=&d_{2A}^2 [M_0+M_1+M_2]+2d_{2A}d_{2B}  [M_0+\frac{1}{2}M_1]\nonumber\\
& &+d_{2BB}^2 M_0 \nonumber \\
&=&(d_{2A}+d_{2B})^2 M_0 +d_{2A} (d_{2A}+d_{2B}) M_1\nonumber\\
& &+d_{2A}^2 M_2
\end{eqnarray}
where the integrals $M_i$ are those defined above in Eq.~(\ref{Equation-P22-M}).
The first expression is, up to simple notational differences, identical to that given 
by \cite{takahashi2008third}.\footnote{An equivalent expression is 
given in \cite{fasiello2016nonlinear}, also in terms of two redshift dependent functions and three integrals which are linear combinations of $M_0$, $M_1$ and $M_2$.}
Making use of the additional relation Eq.~\eqref{relation-d2AB} 
we have derived above, this simplifies to

\begin{equation}\label{PS-22-general}
P_{22}=M_0+d_{2A} M_1 + d_{2A}^2 M_2,
\end{equation}
i.e. to a form evidently equivalent to that in the gEdS model 
Eq.~(\ref{P22-gEdS}) when  we parametrize $d_{2A}$ by $\alpha_{2}$ 
or $\gamma_2$ as defined in
the previous section. 
Defining now 
\begin{equation}
\Delta P_{22}=P_{22}-P_{22}^{EdS},
\end{equation}
where $P_{22}^{EdS}$ is the result in the separable EdS approximation
\begin{equation}
P_{22}^{EdS}=M_0+\frac{5}{7}M_1 + \frac{25}{49} M_2,
\end{equation}
the cosmology dependent correction $\Delta P_{22}$ is expressed 
explicitly as a function of $\gamma_2(a)$ by
\begin{equation}\label{Delta_P22}
\Delta P_{22}=\frac{2}{49}\gamma_2 (M_1+\frac{10}{7}M_2) +(\frac{2}{49} \gamma_2)^2 M_2.
\end{equation}
Recalling our discussion below we see that this correction is thus given in terms just of
integrals that are explicitly infra-red safe, which is not the case if we do not 
make use of the relation Eq.~\eqref{relation-d2AB}. 

\begin{table*}
\caption{A summary of the calculation of the three ``effective growth rates''  parametrizing 
the second and third-order density kernels using the different approximations we have discussed and also the exact 
calculation.}

\begin{ruledtabular}
\begin{tabular}{ccccc} 
 Approximation &   $\alpha_2$ & $\alpha_{3AA}$ & $\alpha_{3AB}$ &   Precision \\ 
 \hline
 EdS & $1$ & $1$ & $1$ & $\lesssim 0.5 
 \%$ for PS at $z=0$ (LCDM) \\
 adiabatic gEdS & $\alpha_1=\frac{d\ln D_1}{d\ln a}$, Eq. \eqref{alpha1} & $\alpha_1$ & $\alpha_1$ & valid for small $|w_0|$ (cf. Figs. \ref{alpha2_and_alpha10}, \ref{alpha_10_alpha_2_alpha_3_small_w}, and \ref{alpha_10_alpha_2_alpha_3_big_w}) \\
gEdS & $\alpha_2$, Eq. \eqref{evolution_gamma2} & $\alpha_2$ & $\alpha_2$& $\lesssim 0.1 \% $ for PS at $z=0$ (LCDM)\\
 exact & $\alpha_2$, Eq. \eqref{evolution_gamma2} & $\alpha_{3AA}$ , Eq.\eqref{evolution_gamma3XX_v1}& $\alpha_{3AB}$, Eq. \eqref{evolution_gamma3XX_v1}   & ---
\end{tabular}
\label{Table-Comparison}
\end{ruledtabular}
\end{table*}

\subsubsection{`13' term}
Starting from Eq.~(\ref{third-order-density-field}), the expression for 
$P_{13}$ can be obtained directly as 
\begin{eqnarray}
2 P_{13}&=&d_{3AA}(N_0+N_3)+d_{3AA}^{\prime}(N_1-N_3)
\nonumber \\ & &+d_{3AB}(N_0+N_3) -d_{3AB}^\prime 
N_3  +d_{3BA}(N_0+N_2)\nonumber \\ & &+d_{3BB}N_0 \nonumber \\
&=&(d_{3AA}+d_{3AB}+d_{3BA}+d_{3BB})N_0 \nonumber \\ & &+d_{3AA}^{\prime} N_1 
+ d_{3BA}N_2 +(d_{3AA}-d_{3AA}^\prime\nonumber \\ & &+d_{3AB}-d_{3AB}^\prime)N_3 , 
\end{eqnarray}
where the integrals $N_0, N_1, N_2$ are those defined by Eq.~(\ref{Equation-P13-N}) and 
Eq.~(\ref{Equation-P13-N-defs}). $N_3$ is an additional integral defined also by 
Eq.~(\ref{Equation-P13-N})  but with   
$n_3=-\frac{4}{3}r^2$, and which, it is simple to verify, is infra-red safe.
The first expression has been written to be easily compared to that 
given by \cite{takahashi2008third} in terms of the six 
functions $d_{3XX}$ and six integrals.\footnote{The equivalent 
expression in \cite{fasiello2016nonlinear} is given in terms of six redshift 
dependent functions and five integrals.}
Making use now of the relations
Eqs.~(\ref{constraints-all1})-(\ref{constraints-all4})
obtained from the EdS (or gEdS) boundary condition, 
we find that the coefficient of $N_3$ vanishes, and 
further that the expression can be simplified to be 
a function of only of $d_{2A}$ and only
one of the initial six functions $d_{3XX}$: 
\begin{equation}\label{PS-13-general}
2 P_{13}=N_0-N_1
+ 2d_{2A}N_1 +d_{3BA}(N_2-N_1),
\end{equation}
Thus, as for $P_{22}$, we obtain an expression involving only
the same three integrals as in the gEdS one-loop PS,
and as in that case, the cosmology-dependent terms 
involve only the two infra-red safe integrals $N_1$
and $N_2$. We note that this latter property is
obtained using the relation Eq.~(\ref{constraints-all3}).
Conversely, without employing this relation,  
which we have shown here follows from the 
condition that the cosmology asymptotes to
EdS (or gEdS) at early times, it is not
possible to recover explicitly the 
infra-red safety of the cosmological correction.


The cosmological correction in $P_{13}$ can thus be taken to depend in general only 
on $\gamma_{2}$, as defined in Eq.~(\ref{def-gamma}), and on just {\it one} additional function. 
It is convenient then to 
introduce the function $\gamma_{3BA}=\gamma_{3}$, defined, following the same treatment as in
Section~\ref{LCDM approximated as an interpolation of gEdS},  as
\begin{equation}
d_{3BA}=\frac{2\alpha_3(1+2\alpha_3)}{(1+6\alpha_3)(1+8\alpha_3)}=\frac{2}{21}\Big[1+\frac{5}{63}\gamma_{3}\Big],
\label{def-gamma-3} 
\end{equation}
so that $\alpha_{3} \approx 1-\gamma_{3}$ for $\gamma_{3}\ll 1$. 
$\gamma_3$ is given in terms of $\gamma_2$ and the two functions 
$\gamma_{3AA}$ and $\gamma_{3BA}$ analysed in the 
Section~\ref{LCDM approximated as an interpolation of gEdS}, by
\begin{eqnarray}
15\gamma_3= 162 \gamma_{2}-196\gamma_{3AA}+49\gamma_{3AB}.
\end{eqnarray}
It is thus a solution to the equation
\begin{equation}   
{\cal {D}}^{(3)} \gamma_{3}=
-\frac{189}{5} \Big(\Omega_{m}-\alpha_{1}^2\Big) +\frac{189}{35}\Big[4\alpha_{1}^2 \gamma_{2} + 2\alpha_1 \frac{d\, \gamma_{2}}{d\ln a}\Big],
\end{equation}
with appropriate boundary conditions.
Defining
\begin{equation}
2\Delta P_{13}=2 P_{13}-2 P_{13}^{EdS},
\end{equation}
where
\begin{equation}
2 P_{13}^{EdS}=N_0+\frac{1}{3}N_1+\frac{2}{21}N_2,
\end{equation}
is the usual result in the separable EdS approximation, 
the cosmological correction is given as a function
of $\gamma_2$ and $\gamma_3$ as  
\begin{equation}\label{Delta_P13}
2\Delta P_{13}= \frac{4}{49} \big[\gamma_2 N_1 +\frac{5}{54}\gamma_3  (N_2-N_1)\big].
\end{equation}

\subsubsection{The cosmological correction to the PS at one loop} 

Combining Eqs.~(\ref{Delta_P22}) and (\ref{Delta_P13}), we get the final expression for the full cosmological (i.e. non-EdS) correction to the one-loop PS as
\begin{eqnarray}\label{Delta_P_one_loop}
\Delta P_{1-loop}&=& \frac{2}{49}\gamma_2 (M_1+\frac{10}{7}M_2) +(\frac{2}{49} \gamma_2)^2 M_2 \nonumber\\
& &+\frac{4}{49} \big[\gamma_2 N_1 +\frac{5}{54}\gamma_3  (N_2-N_1)\big].
\end{eqnarray}
\begin{figure*}
\includegraphics[width=7.5cm,height=7.5cm]{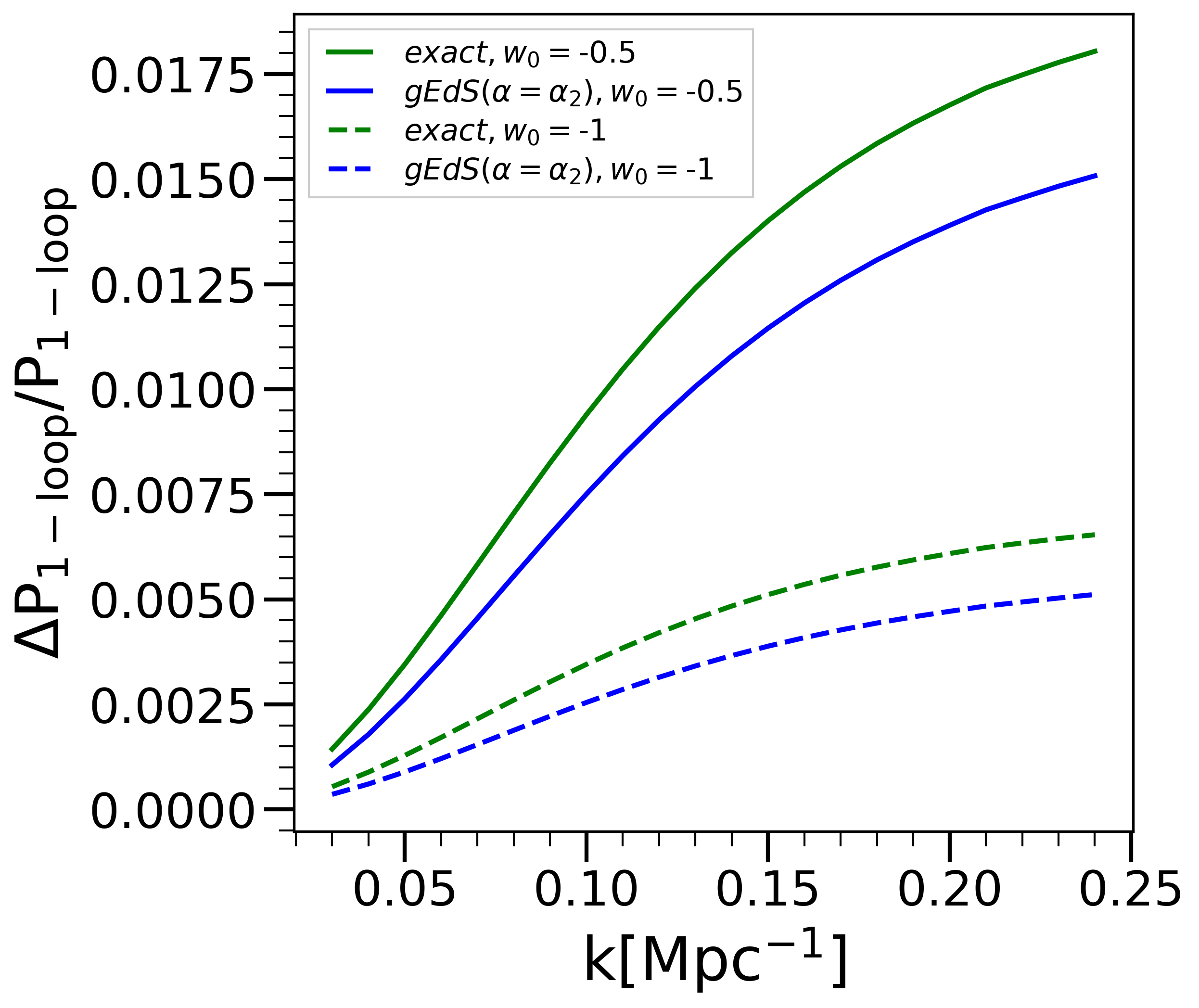}\includegraphics[width=7.5cm,height=7.5cm]{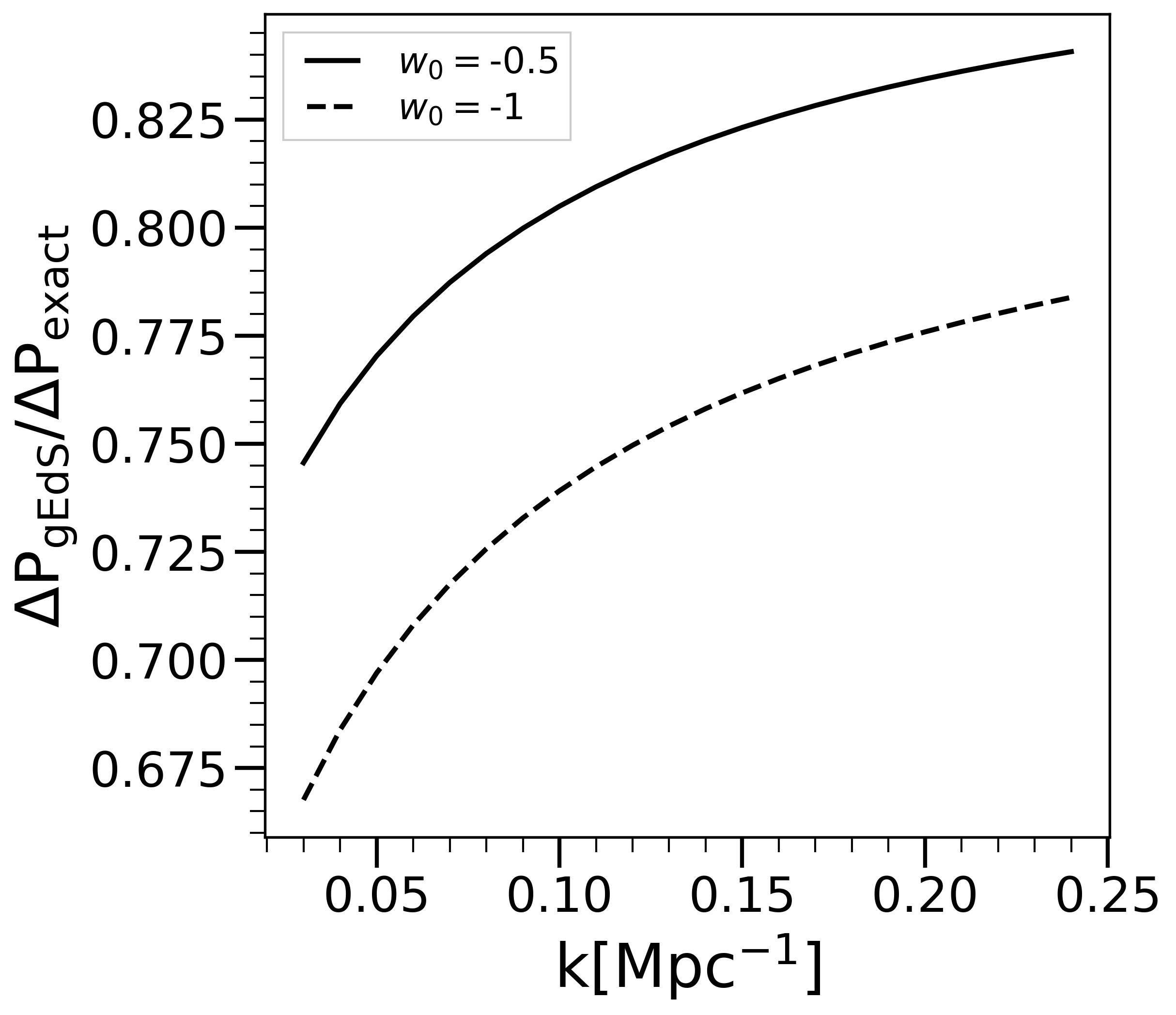}
\caption{Left panel: The one loop cosmological (non-EdS)
correction $\Delta P_{1-loop}$ as a fraction of $P_{1-loop}$, 
the full one-loop  PS in EdS, at z=0, for $w_{0}=-1$ and $w_0=-0.5$ 
(and a typical standard like cosmological PS, see text) calculated 
using the exact expression Eq.~(\ref{Delta_P_one_loop}), and also using the exact 
gEdS kernels with $\alpha$ replaced by the single effective growth 
exponent $\alpha_2(z)$. Right panel: ratio of the gEdS approximation to 
the exact result. 
}
\label{Delta_PS_one_loop}
\end{figure*}
The left panel in Fig.~\ref{Delta_PS_one_loop} shows 
$\Delta P_{1-loop}$ relative to the full one loop 
PS in EdS, at $z=0$, in a standard cosmological
model (specifically, with the cosmological parameters of \cite{ade2014planck} and calculated using the
approximation to the transfer function of \cite{eisenstein1998baryonic}).
The model labeled by $w_0=-0.5$ is identical but for this parameter.
As previously documented in the literature (e.g. \cite{takahashi2008third}) the
corrections are at the sub percent level in the former case but increase
as $|w_0|$ decreases. The reason for this behaviour is very evident in light of our analysis: a smaller $|w|$ is closer to the adiabatic  limit, which corresponds to an 
effective $\alpha$ closer to the instantaneous linear growth 
exponent $\alpha_1=d \log D_1/d \log a$ (i.e. to a larger value of $|\gamma_2|$ and $|\gamma_3|$.)
Also shown are the approximation to $\Delta P_{1-loop}$ obtained  
by directly interpolating with a gEdS model i.e., by replacing $\alpha$ by $\alpha_2(z)$ 
in Eqs.~(\ref{P22-gEdS}) and (\ref{P13-gEdS}) (and subtracting the EdS result from their sum).
The right panel of Fig.~\ref{Delta_PS_one_loop} shows the ratio of this approximation to 
the exact result.
As might be anticipated from the 
expression Eq.~(\ref{Delta_P_one_loop}) --- which depends on the function
$\gamma_3$ only in one term with a small pre-factor --- we find that
interpolation provides a good approximation (within about $25 \% $).
In practice it is not significantly more difficult to calculate the exact result,
but the result shows that for understanding the origin of the smallness of
the calculated correction, the approximation to gEdS is very instructive.
We underline, however, that the adiabatic approximation $\alpha_2 \approx \alpha_{10}$
is not valid, so this is not an adiabatic interpolation in the sense we have defined.
The effective growth exponent $\alpha_2$ can be roughly thought of a time-averaged behaviour 
of the exact linear growth exponent $\alpha_1$, which is itself a good approximation
to $\alpha_2$ for small $|w|$.

To summarize we present in Table \ref{Table-Comparison} a synthesis of the different 
approximations for the calculation of density perturbations up to third order, in cosmological models
with a self-gravitating matter component and a smooth time-dependent component 
(given through an effective equation of state $p=w(a)\rho$).
To parametrize the temporal evolution of the kernels we have defined the three functions $\alpha_2$, $\alpha_{3AA}$ and $\alpha_{3AB}$ (and also equivalent functions $\gamma_2$, $\gamma_{3AA}$ and $\gamma_{3AB}$). 
The table indicates how each of these three functions is calculated in each of the three approximations 
we have discussed for the exact result.

\section{Discussion and conclusions}

We have derived in detail the kernels for both standard EPT and LPT in a family of EdS models, characterized by their constant linear growth rate $\alpha$. While such results for this family of models are implicit in previous 
treatments of perturbation theory in the FLRW cosmologies (e.g. \cite{takahashi2008third}), or in a broader class of cosmologies  (e.g. \cite{fasiello2016nonlinear,fasiello2022perturbation}), the simple analytical results in this case which generalize those of the standard EdS model have not (to our knowledge) been discussed in the literature previously. For the one-loop power spectrum in EFT we have used these kernels to show that the expected infra-red convergence properties of the corrections relative to  a standard EdS model are obtained, without the need for cancellation between the two contributing integrals.
Further we have noted that the coefficients of ultra-violet divergent terms have a strong dependence on $\alpha$, and the leading divergence can even change sign (and is zero at a certain value). 

In the second part of the paper we have shown 
these analytical results for the kernels in these models can be exploited to obtain a simplified formulation of the calculation of the cosmology-dependent corrections to the usual separable EdS approximation in any standard type non-EdS cosmology, and in particular in LCDM-like cosmologies. We do so by parametrizing the relevant time-dependent functions characterizing  this cosmology dependence as effective growth rates in gEdS models. Assuming only that the cosmology at asymptotically early times is gEdS, second-order corrections are parametrized in terms of one such exponent, and third-order corrections in terms of it and two more such exponents.  Thus, for example, the bispectrum calculated at leading non-trivial order in perturbation theory, depending on the second-order perturbation, is, at any time, exactly equal to that in a gEdS model.
For the power spectrum we show the results on the infra-red convergence behaviour of the gEdS model generalize, and
we derive explicit expressions for the corrections to the EdS model expressed in terms of infra-red safe integrals. This expression turns out to depend on only one of the two effective exponents needed at the third order. Nevertheless
the PS is in fact given to a relatively good approximation (up to $\sim 25\%$) by that obtained 
with the analytic gEdS kernels on replacing the fixed growth rate $\alpha$ of gEdS by the 
single effective exponent $\alpha_2(z)$ defining the exact evolution at second order. Our exact expression 
is much simplified compared to previous (equivalent) expressions derived in \cite{takahashi2008third,fasiello2022perturbation} which involve six or eight redshift-dependent functions and do not
explicitly recover the infra-red convergent properties as we have done.  

Our formulation of the cosmological corrections in this way makes it simple to understand why they are so very small in LCDM-like models (typically $< 1\%$ even at $z=0$). The effective growth exponents we introduce to map the time-dependent kernels to those of the gEdS models obey equations in which one can clearly identify an adiabatic limit, of a dark energy component scaling like matter (i.e. $w=0$) in which these exponents converge to the instantaneous logarithmic linear growth rate. The more rapid evolution of dark energy at low redshift acts like a damping of this evolution, leading to the effective exponent much closer to its initial EdS value. Combined with 
the very weak $\alpha$ dependence shown by the analytical expression for the gEdS kernels, the corrections turn out to be as small as they are.
Only for models accelerating much faster than 
$LCDM$ at $z=0$ can such corrections be comparable to the EdS 
approximated loop corrections, and even in the asymptotic limit that 
$\Omega_m \rightarrow 0$ they remain finite and at most of the order unity.

In a forthcoming paper, we will exploit the results for the kernels in gEdS to obtain exact analytical solutions for the PS in what we will call ``generalized scale-free models'' i.e. for a PS which is a simple power law and the cosmology is gEdS \cite{benhaiem2014self, dbenhaiem_PhD}. These models share the property of self-similarity of usual scale-free models (with a standard EdS expansion law). This makes them ideal for numerical testing for the accuracy of the analytical predictions, as the property of self-similarity allows the extraction of highly accurate results from simulations (see \cite{joyce2021quantifying, maleubre2022accuracy}).
In further work we plan also to explore the ultra-violet divergences in perturbation theory in these models, and to examine their regularization by methods such as effective field theory. This may, we anticipate, provide insights into the application of such techniques in standard cosmologies.
It would also be interesting to extend the analysis we have presented here to determine analogous formulations (in terms of effective growth rate functions) of analyses previously reported in the literature for cosmological corrections, both for other statistics 
(two-point velocity statistics, pairwise velocities, trispectrum etc.) at one-loop and two-loop, and also
for a broader range of cosmological models. In particular
it would be interesting to explore such a 
formulation of perturbation theory with 
a massive neutrino component including 
its perturbations (see e.g. 
\cite{ aviles2020lagrangian, aviles2021clustering,garny2021loop,Garny2022Twoloop}). In this case the kernels
have also a scale dependence that might be
expected to be conveniently formulated
in terms of (scale-dependent) effective 
growth exponents analogous to those 
we have employed
here.

\begin{acknowledgments}
A.P. is supported by the Indonesia Endowment Fund for Education (LPDP). The numerical calculations in this research 
used the python packages Matplotlib   \cite{hunter2007matplotlib}, NumPy \cite{harris2020array}, and SciPy \cite{virtanen2020scipy}. 

\end{acknowledgments}

\begin{widetext}
\appendix

\section{PT kernels at third-order in gEdS}\label{third_order_kernels_derivation}
  
\subsection{EPT}
Eq.~(\ref{equation-second-order-density}) at third order in $\delta^{(1)}$ gives  
\begin{eqnarray}\label{specific-second-order-density}
    \mathcal{H}^{2}\Big\{-a^{2}\partial_{a}^{2}-\frac{3}{2}a\partial_{a}&+&\frac{3}{2\kappa^{2}}\Big\}\delta^{(3)}(\textit{\textbf{k}},a)=S^{(3)}_{\Tilde{\beta}}(\textit{\textbf{k}},a)
    -\mathcal{H}\partial_{a}(a S^{(3)}_{\Tilde{\alpha}}(\textit{\textbf{k}},a)),
   \end{eqnarray} 
where
\begin{eqnarray}\label{S_alpha_third_order}
        S_{\Tilde{\alpha}}^{(3)}(\textit{\textbf{k}},a)=-\int \frac{\text{d}^{3}q}{(2\pi)^{3}}\Tilde{\alpha}(\textit{\textbf{q}},\textit{\textbf{k}}-\textit{\textbf{q}},a)\theta^{(1)}(\textit{\textbf{q}},a)\delta^{(2)}(\textit{\textbf{k}}-\textit{\textbf{q}},a)-\int \frac{\text{d}^{3}q}{(2\pi)^{3}}\Tilde{\alpha}(\textit{\textbf{q}},\textit{\textbf{k}}-\textit{\textbf{q}},a)\theta^{(2)}(\textit{\textbf{q}},a)\delta^{(1)}(\textit{\textbf{k}}-\textit{\textbf{q}},a),
   \end{eqnarray} 
\begin{eqnarray}\label{S_beta_third_order}
        S_{\Tilde{\beta}}^{(3)}(\textit{\textbf{k}},a)=-\int \frac{\text{d}^{3}q}{(2\pi)^{3}}\Tilde{\beta}(\textit{\textbf{q}},\textit{\textbf{k}}-\textit{\textbf{q}},a)\theta^{(1)}(\textit{\textbf{q}},a)\theta^{(2)}(\textit{\textbf{k}}-\textit{\textbf{q}},a)-\int \frac{\text{d}^{3}q}{(2\pi)^{3}}\Tilde{\beta}(\textit{\textbf{q}},\textit{\textbf{k}}-\textit{\textbf{q}},a)\theta^{(2)}(\textit{\textbf{q}},a)\theta^{(1)}(\textit{\textbf{k}}-\textit{\textbf{q}},a).
    \end{eqnarray} 
We can write Eq. (\ref{S_alpha_third_order})  as
  \begin{equation}\label{S_alpha_third_order5}
    \begin{split}
        S^{(3)}_{\Tilde{\alpha}}(\textit{\textbf{k}},a)=\mathcal{H}f D^{3}(a) (A_{1}+A_{2}),
    \end{split}
   \end{equation}
where
\begin{eqnarray}
A_1&=&\int \frac{\text{d}^{3}q}{(2\pi)^{3}} \frac{\text{d}^{3}q_{1}}{(2\pi)^{3}}  \frac{\text{d}^{3}q_{2}}{(2\pi)^{3}}
       \delta^{(1)}(\textit{\textbf{q}})\delta^{(1)}(\textit{\textbf{q}}_{1})\delta^{(1)}(\textit{\textbf{q}}_{2})\Tilde{\alpha}(\textit{\textbf{q}},\textit{\textbf{k}}-\textit{\textbf{q}})\textit{F}_{2}(\textit{\textbf{q}}_{1},\textit{\textbf{q}}_{2})(2\pi)^{3}\delta^{(D)}(\textit{\textbf{k}}-\textit{\textbf{q}}-\textit{\textbf{q}}_{1}-\textit{\textbf{q}}_{2}),\\
       A_2&=&\int \frac{\text{d}^{3}q}{(2\pi)^{3}} \frac{\text{d}^{3}q_{1}}{(2\pi)^{3}}  \frac{\text{d}^{3}q_{2}}{(2\pi)^{3}}
       \delta^{(1)}(\textit{\textbf{q}})\delta^{(1)}(\textit{\textbf{q}}_{1})\delta^{(1)}(\textit{\textbf{q}}_{2})\Tilde{\alpha}(\textit{\textbf{q}},\textit{\textbf{k}}-\textit{\textbf{q}})\textit{G}_{2}(\textit{\textbf{q}}_{1},\textit{\textbf{q}}_{2})(2\pi)^{3}\delta^{(D)}(\textit{\textbf{q}}-\textit{\textbf{q}}_{1}-\textit{\textbf{q}}_{2}).
\end{eqnarray}
and Eq. (\ref{S_beta_third_order}) as
    \begin{equation}\label{source1}
    \begin{split}
        S^{(3)}_{\Tilde{\beta}}(\textit{\textbf{k}},a)=-\mathcal{H}^{2}f^{2} D^{3}(a) (B_{1}+B_{2}), 
    \end{split}
   \end{equation}
   where
    \begin{eqnarray}\label{S_alpha_third_order2}
       B_1&=&\int \frac{\text{d}^{3}q}{(2\pi)^{3}} \frac{\text{d}^{3}q_{1}}{(2\pi)^{3}}  \frac{\text{d}^{3}q_{2}}{(2\pi)^{3}}
       \delta^{(1)}(\textit{\textbf{q}})\delta^{(1)}(\textit{\textbf{q}}_{1})\delta^{(1)}(\textit{\textbf{q}}_{2})\Tilde{\beta}(\textit{\textbf{q}},\textit{\textbf{k}}-\textit{\textbf{q}})\textit{G}_{2}(\textit{\textbf{q}}_{1},\textit{\textbf{q}}_{2})(2\pi)^{3}\delta^{(D)}(\textit{\textbf{k}}-\textit{\textbf{q}}-\textit{\textbf{q}}_{1}-\textit{\textbf{q}}_{2}),\\
       B_2&=&\int \frac{\text{d}^{3}q}{(2\pi)^{3}} \frac{\text{d}^{3}q_{1}}{(2\pi)^{3}} \frac{\text{d}^{3}q_{2}}{(2\pi)^{3}}
       \delta^{(1)}(\textit{\textbf{q}})\delta^{(1)}(\textit{\textbf{q}}_{1})\delta^{(1)}(\textit{\textbf{q}}_{2})\Tilde{\beta}(\textit{\textbf{q}},\textit{\textbf{k}}-\textit{\textbf{q}})\textit{G}_{2}(\textit{\textbf{q}}_{1},\textit{\textbf{q}}_{2})(2\pi)^{3}\delta^{(D)}(\textit{\textbf{q}}-\textit{\textbf{q}}_{1}-\textit{\textbf{q}}_{2}).
   \end{eqnarray} 
The right-hand side of Eq.~\eqref{specific-second-order-density} can then be written as 
 \begin{eqnarray}
        S^{(3)}_{\Tilde{\beta}}-\mathcal{H}\partial_{a}\big(a S^{(3)}_{\Tilde{\alpha}})=-\mathcal{H}^{2}\alpha^{2} a^{3\alpha} (B_{1}+B_{2})-\mathcal{H}^{2}a^{3\alpha}\Big[\frac{1}{2}\alpha+3\alpha^2 \Big](A_1+A_2),\nonumber\\
\end{eqnarray}
from which we obtain 
 \begin{eqnarray}
        \delta^{(3)}(\textbf{k})=\frac{1}{2(8\alpha+1)} \bigg[2\alpha(B_{1}+B_{2})+(1+6\alpha)(A_1+A_2)\bigg],\nonumber\\
\end{eqnarray}
which corresponds to 
 \begin{eqnarray} 
      \textit{F}_{3} (\textit{\textbf{q}}_{1},\textit{\textbf{q}}_{2},\textit{\textbf{q}}_{3})& =&\frac{1}{2(8\alpha+1)}\bigg[(6\alpha+1)\Tilde{\alpha}(\textit{\textbf{q}}_{1},\textit{\textbf{q}}_{2}+\textit{\textbf{q}}_{3})\times\textit{F}_{2}(\textit{\textbf{q}}_{2},\textit{\textbf{q}}_{3})+(2\alpha)\Tilde{\beta}(\textit{\textbf{q}}_{1},\textit{\textbf{q}}_{2}+\textit{\textbf{q}}_{3})\textit{G}_{2}(\textit{\textbf{q}}_{2},\textit{\textbf{q}}_{3})\nonumber\\
      & &+\big[(6\alpha+1)\Tilde{\alpha}(\textit{\textbf{q}}_{1}+\textit{\textbf{q}}_{2},\textit{\textbf{q}}_{3})+(2\alpha)\Tilde{\beta}(\textit{\textbf{q}}_{1}+\textit{\textbf{q}}_{2},\textit{\textbf{q}}_{3})\big]\textit{G}_{2}(\textit{\textbf{q}}_{1},\textit{\textbf{q}}_{2}) \bigg].\nonumber\\
 \end{eqnarray} 
  
 \subsection{LPT}
  \label{Appendix-Lagrangian perturbation theory}
The left-hand side of Eq.~\eqref{equation_displacement}  at third order 
can be written
    \begin{eqnarray}
        & &\frac{(18\alpha^{2}+3\alpha)}{2}\Psi_{i,i}^{(3)}-(4\alpha^{2}+\alpha)\Psi_{i,j}^{(1)}\Psi_{i,j}^{(2)}-\frac{(2\alpha^{2}+\alpha)}{2}\Psi_{i,j}^{(2)}\Psi_{i,j}^{(1)}+(4\alpha^{2}+\alpha)\Psi_{i,i}^{(1)}\Psi_{j,j}^{(2)}\nonumber\\
        & &-\frac{(2\alpha^{2}+\alpha)}{2}\Psi_{i,i}^{(1)}\Psi_{i,j}^{(1)}\Psi_{i,j}^{(1)}+\frac{(2\alpha^{2}+\alpha)}{2}\Psi_{i,i}^{(2)}\Psi_{i,j}^{(1)}+\frac{(2\alpha^{2}+\alpha)}{4}\Big\{\Psi_{i,i}^{(1)}\Psi_{j,j}^{(1)}\Psi_{k,k}^{(1)}-\Psi_{i,i}^{(1)}\Psi_{j,k}^{(1)}\Psi_{k,j}^{(1)}\Big\}\nonumber\\
        & &+\frac{(2\alpha^{2}+\alpha)}{2}\Psi_{i,l}^{(1)}\Psi_{l,j}^{(1)}\Psi_{i,j}^{(1)},\label{eq:43}
\end{eqnarray}
and the right-hand side as
\begin{eqnarray}
      & &\frac{(2\alpha^{2}+\alpha)}{2}\Psi_{i,i}^{(3)}+\frac{(2\alpha^{2}+\alpha)}{2}\big\{\Psi_{i,i}^{(2)}\Psi_{j,j}^{(1)}-\Psi_{i,j}^{(2)}\Psi_{j,i}^{(1)}\big\}+\frac{(2\alpha^{2}+\alpha)}{2}\Big\{\frac{1}{6}\Psi_{i,i}^{(1)}\Psi_{j,j}^{(1)}\Psi_{k,k}^{(1)}\nonumber\\
       & &-\frac{1}{2}\Psi_{i,i}^{(1)}\Psi_{j,k}^{(1)}\Psi_{k,j}^{(1)}+\frac{1}{3}\Psi_{i,l}^{(1)}\Psi_{l,j}^{(1)}\Psi_{i,j}^{(1)}\Big\}\label{eq:44}.
\end{eqnarray}
Collecting up the terms we have
\begin{eqnarray}
       (8\alpha+1)\Psi_{i,i}^{(3)}&=&-(4\alpha+1)\Psi_{i,i}^{(2)}\Psi_{j,j}^{(1)}+(4\alpha+1)\Psi_{i,j}^{(2)}\Psi_{j,i}^{(1)}+\frac{(2\alpha+1)}{2}\Psi_{i,i}^{(1)}\Psi_{i,j}^{(1)}\Psi_{j,i}^{(1)}-\frac{(2\alpha+1)}{6}\Psi_{i,i}^{(1)}\Psi_{j,j}^{(1)}\Psi_{k,k}^{(1)}\nonumber\\
       & &-\frac{(2\alpha+1)}{3}\Psi_{i,k}^{(1)}\Psi_{k,j}^{(1)}\Psi_{j,i}^{(1)},
\end{eqnarray}
which gives Eq.~\eqref{Third-order-LPT}.

\section{Ultra-violet divergences in one-loop PS for gEdS}
\label{Appendix-asymptotic behavior of the one-loop power spectrum}
We detail here a little more the analysis leading to the expressions
Eq.~(\ref{eq:asymptotic_P22_big_q}) and Eq.~(\ref{eq:asymptotic_P13_big_q})
which allows us to infer the ultraviolet convergence properties of
these integrals i.e. the asymptotic large $k$ behaviour of $P_{\rm lin}(k)$
for which they converge. 

For the $P_{22}$ the leading divergence is obtained simply by replacing 
$P_{\rm lin} (|\textbf{k}-\textbf{q}|)$ by  $P_{\rm lin}(q)$ and 
taking the leading term in the series expansion in $k/q$
around $(k/q)=0$: 
\begin{eqnarray}\label{Expansion_kernel_F2_k_to_zero}
|F_{2}^{(s)}(\textbf{k}-\textbf{q},\textbf{q})|^{2}&=&\frac{1}{4}\frac{k^4}{q^4}+\Big(\frac{4\alpha+1}{6\alpha+1}\Big)\Big[(\mu^2-1)\frac{k^4}{q^4}\Big]+\Big(\frac{4\alpha+1}{6\alpha+1}\Big)^2 \Big[(\mu^2-1)^2 \frac{k^4}{q^4}\Big]+\dots\nonumber\\
&=&\Big(\frac{2 \alpha -(8 \alpha +2) \mu^2+1}{2(6 \alpha +1)}\Big)^2 \frac{k^4}{q^4} + \cdots\nonumber\\
\end{eqnarray}
where $\mu=\frac{\textbf{k}\cdot\textbf{q}}{k q}$ is the angular variable.
Using Eqs.~(\ref{equation-P22}) and (\ref{Expansion_kernel_F2_k_to_zero}) we then 
obtain, after integrating $\mu$, Eq.~(\ref{eq:asymptotic_P22_big_q}).

Likewise expanding $\int_{-1}^{1}d\mu F_{3}^{(s)}(\textbf{k},\textbf{q},-\textbf{q})$ 
in a power series in $k/q$ around $(k/q)=0$ we obtain 
\begin{eqnarray}\label{Asymptotic_kernel_F3_k_zero}
\lim_{k/q \to 0}\int_{-1}^{1}d\mu F_{3}^{(s)}(\textbf{k},\textbf{q},-\textbf{q})&=&-\frac{1}{9}\frac{k^2}{q^2}+\Big(\frac{2 \alpha +1}{8 \alpha +1}\Big)\Big[-\frac{4}{105}\frac{k^4}{q^4}+\times\frac{4}{15}\frac{k^2}{q^2}\Big]+\Big(\frac{2 \alpha +1}{8 \alpha +1}\Big)\Big(\frac{2 \alpha}{6 \alpha +1}\Big)\Big[\frac{4}{15}\frac{k^4}{q^4}-\frac{4}{9}\frac{k^2}{q^2} \Big]+\dots\nonumber\\
&=&\frac{7-14\alpha-176\alpha^{2}}{45(1+6\alpha)(1+8\alpha)}\frac{k^{2}}{q^{2}}+\frac{4(8\alpha-1)(1+2\alpha)}{105(1+6\alpha)(1+8\alpha)}\frac{k^{4}}{q^{4}}+\dots
\end{eqnarray}
which leads directly to Eq.~(\ref{eq:asymptotic_P13_big_q}).

\section{Summary of calculation of cosmology dependent corrections to the EdS approximation for the one loop PS in standard perturbation theory}

We have shown that this correction is given by 
\begin{eqnarray}\label{Delta_P_one_loop-appendix}
\Delta P_{1-loop}= \frac{2}{49}\gamma_2 (M_1+\frac{10}{7}M_2) +(\frac{2}{49} \gamma_2)^2 M_2 +\frac{4}{49} \big[\gamma_2 N_1 +\frac{5}{54}\gamma_3  (N_2-N_1)\big],
\end{eqnarray}
where the four integrals are given in Eqs.~\eqref{Equation-P22-M} and \eqref{Equation-P13-N}.
These expressions are valid for a generic FLRW cosmology with a clustering matter component and a smooth component that scales at asymptotically early times as matter i.e. with $w\rightarrow0$ as $a \rightarrow 0$ (where 
$w=\frac{1}{3}\frac{d\Omega_m (a)}{d\ln a}$).  
The two functions $\gamma_2(a)$ and $\gamma_3(a)$
are obtained by solving the coupled equations:
\begin{eqnarray}\label{eq-alpha1}
\frac{d\alpha_{1}}{d\ln a}&=&\frac{3}{2}w \alpha_{1}-\alpha_{1}^2
   - \frac{1}{2} \alpha_{1}
    +\frac{3}{2}\Omega_{m},\\
     \frac{d^2 \gamma_{2}}{d\ln a^2} &=&-\Big[\frac{1}{2}(1-3w) + 4\alpha_{1}\Big]\frac{d \gamma_{2}}{d\ln a} -\Big(2\alpha_{1}^2+\frac{3}{2}\Omega_{m}\Big) \gamma_2+ \frac{21}{2}(\Omega_{m}-\alpha_{1}^2)\label{eq-gamma2},\\
      \frac{d^2 \gamma_{3}}{d\ln a^{2}}&=&-\Big[\frac{1}{2}(1-3w)+6\alpha_{1}\Big]\frac{d\gamma_{3}}{d\ln a}-3(2\alpha_{1}^2+\Omega_{m})\gamma_{3} 
    -\frac{189}{5} \Big(\Omega_{m}-\alpha_{1}^2\Big)+\frac{189}{35}\Big[4\alpha_{1}^2 \gamma_{2} + 2\alpha_1 \frac{d\gamma_{2}}{d\ln a}\Big],\label{eq-gamma3}
\end{eqnarray}
subject to the boundary conditions, given at $a=0$, by 
\begin{equation}
\alpha_1(0)=\alpha_{2}(0)=\alpha_3(0)=\alpha_{10}(0),\, \frac{d\alpha_2}{da} (0)=\frac{d\alpha_3}{da} (0)=0,
\end{equation}
or, more explicitly,
\begin{eqnarray}
 \gamma_2(0)&=&7\frac{1-\alpha_{10}(0)}{1+6 \alpha_{10}(0)}, \nonumber\\ 
 \gamma_3(0)&=&\frac{63}{5}\frac{(1-\alpha_{10}(0))(6 \alpha_{10}(0)-1)}{(1+6 \alpha_{10}(0))1+8\alpha_{10}(0))},
\end{eqnarray}
and 
\begin{equation}
\frac{d\gamma_2}{da} (0)=0\,\,,\frac{d\gamma_3}{da} (0)=0,
\end{equation}
and where $\alpha_{10}(0)$ is given by Eq.~(\ref{alpha-adiabatic}) with
$\Omega_m$ given by the asymptotic (constant) fraction of clustering 
matter at high redshift.
Given that $\alpha_1=\alpha_{2}=\alpha_3(0)=\alpha_{10}(a)$ is the solution 
at leading order in $w$, we can integrate in practice from $0<a_i\ll 1$ 
assuming exactly the initial conditions given by this solution.
Note that in a standard EdS cosmology we have of course 
$\alpha_{10}(0)=1$, and the case  $\alpha_{10}(0)\neq 1$ corresponds to 
the more general gEdS case where part of the matter component at 
high redshift is smooth (e.g. a massive neutrino component or matter-like 
dark energy component).

\section{Detailed comparison with results of \cite{takahashi2008third}}\label{Appendix-Takehashi}
Expressions for the cosmological corrections,  at second 
and third order in EPT, and for the one-loop PS (as in the previous appendix), have been given in \cite{takahashi2008third}, in terms of six time-dependent functions. Further this paper provides phenomenological fits for the parameter dependence in the case of a smooth component with a constant equation of state. We make explicit here how these six functions are related to one another by the three additional constraints we have derived 
(which reduces the number of independent functions to three, of which only two 
are required to calculate the PS at one loop). We also check that there is indeed good numerical agreement between our results and those of \cite{takahashi2008third}. 

\subsection{Second-order growth factor}
\cite{takahashi2008third} gives results for the second-order growth factor in terms of two numerically fitted functions as

\begin{eqnarray}
\label{correction_Takahasi_2A}
T_{2A}&=&\frac{D_{2A}}{D_{1}^2}-1 
=|\ln{\Omega_m}|\Big(\frac{5.54\times 10^{-3}}{|w_0|}-\frac{3.40\times 10^{-3}}{\sqrt{|w_0|}}\Big),\\
\label{correction_Takahasi_2B}
T_{2B}&=&\frac{D_{2B}}{D_{1}^2}-1  =|\ln{\Omega_m}|\Big(-\frac{1.384\times 10^{-2}}{|w_0|}+\frac{8.50\times 10^{-3}}{\sqrt{|w_0|}}\Big).
\end{eqnarray}
These are related (cf. Eqs. \eqref{def-gamma}) to our effective exponent $\gamma_2$
by 
\begin{equation}
T_{2A}=\frac{2}{35}\gamma_{2},\qquad T_{2B}=-\frac{1}{7}\gamma_{2},
\end{equation}
and thus $T_{2A}=-(2/5)T_{2B}$. Comparing the coefficients we see that this relation is indeed verified within the level of numerical precision quoted by \cite{takahashi2008third}.\footnote{What we have denoted by $w_0$ here corresponds to $w$ in \cite{takahashi2008third}.} 

\begin{figure}[t]
\includegraphics[width=7cm, height=7cm]{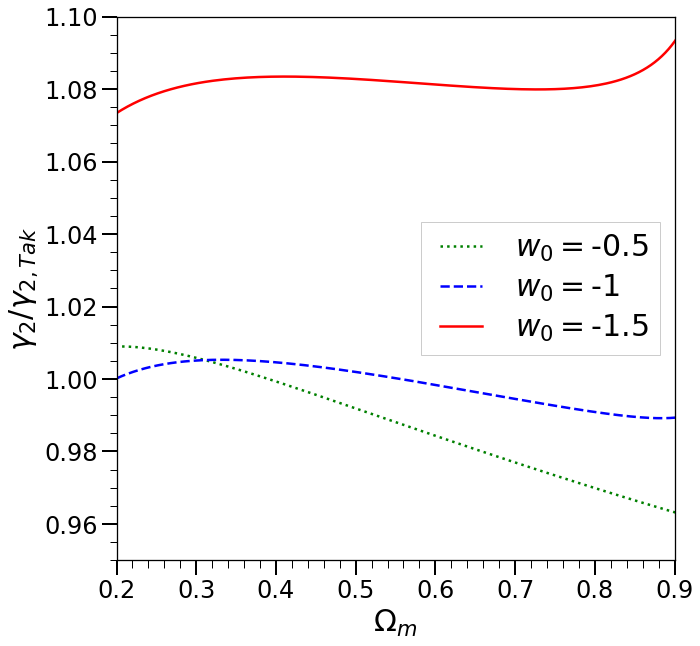}
\caption{The ratio of our numerical solutions for $\gamma_2$ to the parametric fit 
of \cite{takahashi2008third} $\gamma_{2,Tak}$ for the same quantity, for 
dark energy with constant equation of state $w_0=-0.5,-1.0$, and $-1.5$.}
\label{comparison_tak_2}
\end{figure}

\begin{figure*}[t]
   \includegraphics[width=7.5cm,height=7.5cm]{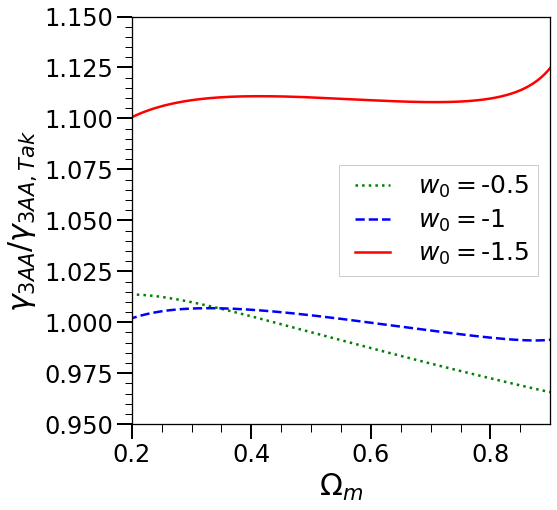}\includegraphics[width=7.5cm,height=7.5cm]{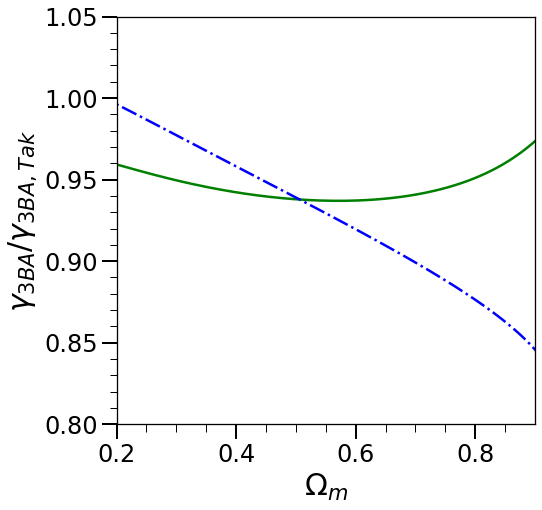}\caption{Left panel: Ratio of the function $\gamma_{3AA}$ given by our direct numerical solution of Eq.~(\ref{evolution_gamma3XX_v0}) to its value inferred from the fits of  \cite{takahashi2008third}, for a smooth component with constant equation of state ($w_{0}=-0.5,-1$ and $-1.5$).
    Right panel: Same comparison, but only for the model with $w_0=-1$,  
    and for $\gamma_{3}=\gamma_{3BA}$ (from numerical solution of Eq.~(\ref{eq-gamma3})). The two different curves correspond to the
    two different possible numerical fits which can be inferred 
    from those of \cite{takahashi2008third}: the full line corresponds to
    that inferred from $T_{3BA}$ directly, and the dashed line to that
    obtained using the linear combination on the right-hand side 
    of Eq.~(\ref{Tak-rel3_2})). The level of accuracy is less good than for $\gamma_2$ and $\gamma_{3AA}$, but quite adequate for calculation notably of the PS at any practically relevant level of precision. }
    \label{comparison-tak-3}
\end{figure*}
Fig.~\ref{comparison_tak_2} shows the ratio between the results we obtain for
$\gamma_2$ by numerical integration of Eqs.(\ref{eq-alpha1})
and (\ref{eq-gamma2}) with the fit for the same quantity 
inferred from \cite{takahashi2008third}: 
\begin{eqnarray}\label{gamma2-Takahashi-fit}
\gamma_{2,Tak}=(35/2)T_{2A}\approx|\ln{\Omega_m}|\Big(\frac{9.7\times 10^{-2}}{|w_0|}-\frac{5.95\times 10^{-2}}{\sqrt{|w_0|}}\Big),\nonumber\\
\end{eqnarray}
We see that our numerical solution for $\gamma_2$ well agree  
very accurately with Eq.~\eqref{gamma2-Takahashi-fit} for $w_0=-0.5,-1$
and well, albeit slightly less so, for $w_0=-1.5$.

\subsection{Third-order growth factor}
At this order \cite{takahashi2008third} gives results in terms of the four functions
\begin{eqnarray}
\label{Tak-fits3}
T_{3AA}=\frac{D_{3AA}}{D_{1}^3}-1&\simeq&  |\ln{\Omega_m}|\Big(\frac{8.21\times 10^{-3}}{|w_0|}-\frac{5.14\times 10^{-3}}{\sqrt{|w_0|}}\Big),\\
T_{3AB}=\frac{D_{3AB}}{D_{1}^3}-1&\simeq& |\ln{\Omega_m}|^{1.5+0.4\ln{|w_0|}}|\Omega_{m}|^{0.7|w_0|}\Big(-\frac{9.16\times 10^{-3}}{|w_0|}+\frac{8.95\times 10^{-3}}{\sqrt{|w_0|}}\Big),\\
T_{3BA}=\frac{D_{3BA}}{D_{1}^3}-1&\simeq& |\ln{\Omega_m}|^{1.06-0.5\ln{|w_0|}}\Big(7.68\times 10^{-3}|w_0|-1.130\times 10^{-2}\sqrt{|w_0|}\Big),\\
T_{3BB}=\frac{D_{3BB}}{D_{1}^3}-1&\simeq& |\ln{\Omega_m}|\Big(-\frac{2.641\times 10^{-2}}{|w_0|}+\frac{1.582\times 10^{-2}}{\sqrt{|w_0|}}\Big),
\end{eqnarray}
and two additional functions $D_{3AA}^\prime$ and $D_{3AB}^\prime$ determined by the relations in Eqs. \eqref{constraints}. 

These are related to the functions we have introduced by 
\begin{eqnarray}
T_{3AA}&=&\frac{4}{45}\gamma_{3AA},\quad T_{3AB}=-\frac{1}{9}\gamma_{3AB}, \nonumber\\
T_{3BA}&=&\frac{5}{63}\gamma_{3BA},\quad T_{3BB}=-\frac{16}{63}\gamma_{3BB}.
\end{eqnarray}

From Eqs.~\eqref{constraint_d3xx}-\eqref{constraint_d3xx_2} it is straightforward to show that our
additional constraints relating to these functions are
\begin{eqnarray}
98\gamma_{3AA}-49 \gamma_{3AB} +15 \gamma_{3BA}-64 \gamma_{3BB}=0,\\
49\gamma_{3AA}+32 \gamma_{3BB} =81 \gamma_{2}.
\end{eqnarray}
Using these we can infer the relations 
\begin{eqnarray}
\label{Tak-rel3_1}
T_{3BB}
&=&\frac{9}{2}T_{2B}+\frac{35}{8}T_{3AA},
\\
\label{Tak-rel3_2}
T_{3BA}
&=&-\big( \frac{4}{3}T_{3BB}+\frac{35}{6}T_{3AA}+\frac{7}{3}T_{3AB}\big).
\end{eqnarray}
For the first relation it is easy to check the accuracy with which 
it is satisfied by the expressions above from \cite{takahashi2008third},  because the  same functional form has been used to fit the functions on both sides: comparing the two fitted coefficients on the left and right-hand side of the relation Eq.~(\ref{Tak-rel3_1}), we find they agree at the sub-percent level. Further to check the numerical accuracy of these fits against ours we plot in the left panel of Fig.~\ref{comparison-tak-3} 
the ratio between our own numerical solution for $\gamma_{3AA}$ and 
$\gamma_{3AA,Tak}=(45/4)T_{3AA}$. The agreement is at a comparable level
to that for $\gamma_2$.

To assess both the agreement of Eq.~(\ref{Tak-rel3_2}) with the fits
of \cite{takahashi2008third} given above, and the accuracy of these fits in describing the exact solutions, we plot in the right panel of 
Fig.~\ref{comparison-tak-3} the ratio of 
our numerical solutions for $\gamma_3=\gamma_{3BA}$ to
$\gamma_{3BA,Tak1}=(63/5)T_{3BA}$ 
and $\gamma_{3BA,Tak2}=(63/5)T_{3BA}^\prime$ respectively,
where $T_{3BA}^\prime$ is the linear combination on the
right-hand side of Eq.~(\ref{Tak-rel3_2}). These curves are
for a model with $w_0=-1$. We can infer that Eq.~(\ref{Tak-rel3_2})
is obeyed by the fitting formulae at the $10\%$ level, and that 
the agreement with our numerical results is slightly 
better if we use the fit for $T_{3BA}$. We have checked carefully
the origin of these small discrepancies and find that they arise
from transients from initial conditions which disappear. 
In practice however, notably for the PS corrections which depend 
only very weakly on $\gamma_3$, the fits provided by 
\cite{takahashi2008third} are accurate enough for any current practical application. 



\section{Comparison  with analysis of \cite{fasiello2022perturbation}}\label{Appendix-comparison-to-Fasiello}
\begin{figure}[b]
    \includegraphics[width=8cm,height=8cm]{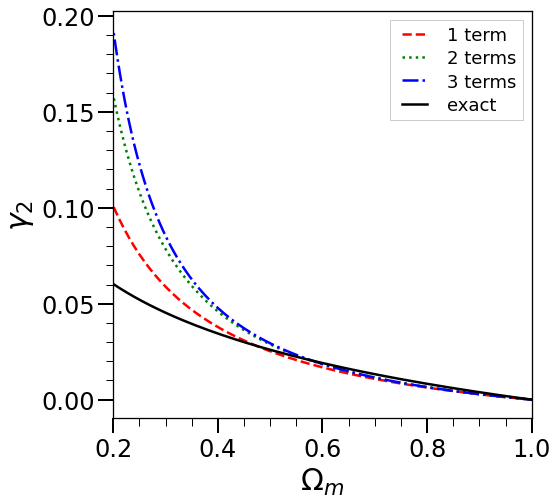}
    \includegraphics[width=8cm,height=8cm]{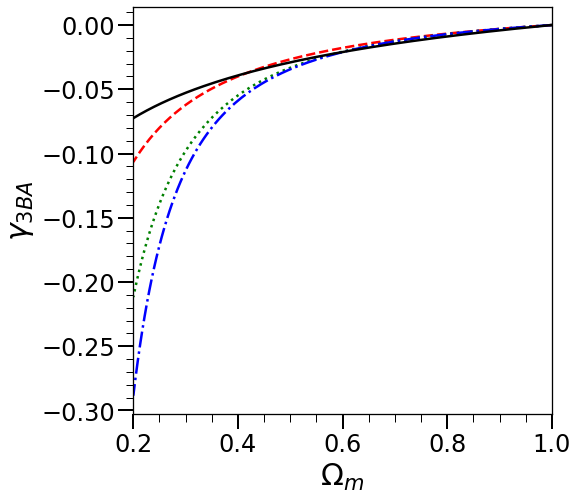}
    \caption{Our numerical solutions for $\gamma_2$,
    and $\gamma_{3}=\gamma_{3BA}$ (solid black lines) in 
    LCDM ($w_0=-1$) compared with the power series expressions in Eqs. \eqref{gamma2-Vlah}-\eqref{gamma3BA-Vlah}
    obtained from \cite{fasiello2022perturbation}.
    The different lines correspond to including terms
    in the power series up to the indicated order.}
   \label{Comparison_to_Vlah}
\end{figure}

In \cite{fasiello2022perturbation} the time dependence of the second order kernel is
parametrized by two functions $\lambda_{2}^{(1)}$ and $\lambda_{2}^{(2)}$, which
are identical to the two functions we have used:
\begin{eqnarray}\label{comparing-second-kernels-to-Vlah}
\lambda_{2}^{(1)}=d_{2A},\quad \lambda_{2}^{(2)}=d_{2B}.
\end{eqnarray}
The six functions used to parametrize the third-order density kernels just differ
by a factor of $2$:
\begin{eqnarray}\label{comparing-third-kernels-to-Vlah}
\lambda_{3}^{(1)}=\frac{1}{2}d_{3AA},\quad\lambda_{3}^{(2)}=\frac{1}{2}d_{3AB},\nonumber\\
\lambda_{3}^{(3)}=\frac{1}{2}d_{3AA}^{\prime},\quad
\lambda_{3}^{(4)}=\frac{1}{2}d_{3AB}^{\prime},\nonumber\\
\lambda_{3}^{(5)}=\frac{1}{2}d_{3BA},\quad
\lambda_{3}^{(6)}=\frac{1}{2}d_{3BB}.
\end{eqnarray}
For the specific case of a $LCDM$ cosmology, \cite{fasiello2022perturbation} provide analytical solutions for these
eight functions (and further ones parametrizing the fourth and fifth order kernels) as power series in the parameter 
$\zeta=\frac{\Omega_{\Lambda 0}}{\Omega_{m 0}} e^{3\eta}$,
where $\eta\equiv \ln D_{+}$ ($D_+$ the linear growth factor normalized to unity at $z=0$, and $\Omega_{\Lambda 0}$ and $\Omega_{m 0}$ are the dark energy and matter fractions at $z=0$, respectively). It is straightforward to verify that the expressions given in Appendix A of \cite{fasiello2022perturbation}, up to third order in $\eta$, satisfy the five relations Eq. \eqref{relation-d2AB} and  
Eqs.~(\ref{constraints-all1})-(\ref{constraints-all4}).

To check the numerical agreement of our results with those of \cite{fasiello2022perturbation} we calculate the functions we have 
used to parametrize the PS and find    
\begin{eqnarray}
\gamma_{2}&=&\frac{7}{2}(7\lambda_{2}^{(1)}-5)=-\frac{7  c_1}{26}\zeta -\frac{2 c_2 }{19}\zeta ^2-\frac{7 c_3}{125}\zeta ^3,\label{gamma2-Vlah}\\
\gamma_{3BA}&=&\frac{63}{5}(21\lambda_{3}^{(5)}-1)=\frac{462   c_1}{1625}\zeta+\frac{51  c_2}{266}\zeta ^2+\frac{2576  c_3}{20625}\zeta ^3\label{gamma3BA-Vlah},
\end{eqnarray}
where 
\begin{equation}
c_1=-\frac{3}{32},\quad c_2=-\frac{141}{4114},\quad c_3=-\frac{9993}{1040842}
\end{equation}
Figure \ref{Comparison_to_Vlah} shows the comparison
between our numerical solutions to the exact equations
for the functions $\gamma_2$ and $\gamma_3=\gamma_{3BA}$
and the power series solutions of 
\cite{fasiello2022perturbation}
 for the same quantities, in LCDM models, up to the
 indicated orders. We observe excellent agreement 
 where the power series solutions appear to 
 converge well. 
 
\section{Comparison with analysis of \cite{bernardeau1993skewness,garny2021loop}}\label{Appendix-comparison-to-Bernardeau}

\cite{bernardeau1993skewness} 
parametrized the time dependence of second order
and third order density kernels in terms of
three functions. Comparing their definitions directly to 
ours (equivalent to those of \cite{takahashi2008third, fasiello2022perturbation})
we obtain 
\begin{eqnarray}
    d_{2A}&=&-\frac{1}{2}+\frac{3}{4}\nu_2,\\
    d_{2B}&=&\frac{3}{2}-\frac{3}{4}\nu_2,\\
    d_{3AA}&=&-\frac{\lambda_3}{2}-\frac{3 \nu_2}{4}+\frac{3 \nu_3}{8}+\frac{1}{2},\\
     d_{3AA}^\prime&=&\frac{\lambda_3}{6}-\frac{3 \nu_2}{4}+\frac{3 \nu_3}{8}+\frac{1}{6},\\
      d_{3AB}&=&\frac{7 \lambda _3}{6}+\frac{3 \nu _2}{4}-\frac{3 \nu _3}{8}+\frac{1}{6},\\
      d_{3AB}^\prime&=&\frac{\lambda _3}{2}+\frac{3 \nu _2}{4}-\frac{3 \nu _3}{8}+\frac{1}{2},\\
      d_{3BA}&=&-\frac{\lambda _3}{6}+\frac{9 \nu _2}{4}-\frac{3 \nu _3}{8}-\frac{13}{6},\\
      d_{3BB}&=&-\frac{\lambda _3}{2}-\frac{9 \nu _2}{4}+\frac{3 \nu _3}{8}+\frac{5}{2}.
\end{eqnarray}
It is straightforward to verify that, using these expressions, the five 
relations Eq. \eqref{relation-d2AB} and Eqs. \eqref{constraints-all1}-\eqref{constraints-all4}
are satisfied. As we have shown in Section \ref{LCDM approximated as an interpolation of gEdS}, this corresponds to assuming that the eight functions 
obey conditions valid for EdS boundary conditions,  
as assumed by \cite{bernardeau1993skewness}.
As we have seen in Section \ref{LCDM approximated as an interpolation of gEdS}, these same conditions are also appropriate for the more general case of gEdS boundary conditions.

The three effective growth rate functions ($\alpha_2$, $\alpha_{3AA}$, and $\alpha_{3AB}$) that 
we have chosen to characterize the second order and third order kernels can be
inferred from 
\begin{eqnarray}
    \nu_2 &=&\frac{2}{3} \left(2 d_{2 A}+1\right),\\
    \nu_3&=&\frac{2}{3} (4 d_{2A}+7 d_{3AA}+3d_{3AB}-2),\\
    \lambda_3 &=&\frac{1}{2} (3 d_{3AA}+3 d_{3AB}-2).
\end{eqnarray}
by substituting the appropriate definitions,  
Eq.~(\ref{def-alpha-eff-2}) and Eqs.~(\ref{def-gammas-3}).
In terms of the functions $\gamma_2$, $\gamma_{3AA}$, and $\gamma_{3AB}$ we have
\begin{eqnarray}
\label{bern-params-ours}
    \nu_2 &=& \frac{2}{147} (119 + 4 \gamma_2),\\
    \nu_3&=&\frac{2}{11907} (1372 \gamma_{3AA}-294 \gamma_{3AB}+648 \gamma_2+21483),\\
    \lambda_3 &=&\frac{1}{54} (4 \gamma_{3AA}-2 \gamma_{3AB}+9)\label{bern-params-ours2}.
\end{eqnarray}

\cite{garny2021loop} give (Appendix A) numerical values for $\nu_2$, $\nu_3$ and
$\lambda_3$ in an LCDM model with $\Omega_m=0.3$ at $z=0$. For this case our numerical
integration gives $\gamma_2=0.0452\dots$, $\gamma_{3AA}=0.0417\dots$ and $\gamma_{3AB}=0.00142\dots$.
Using Eqs.~(\ref{bern-params-ours})-(\ref{bern-params-ours2}) we find numerical agreement up to at least four significant 
decimal places.


We note also that our result,  that the one loop PS can be written in terms of $d_{2A}$ and $d_{3BA}$ only, implies that it can equivalently be written only in terms of $\nu_2$ and the linear combination $(4\lambda_3+9\nu_3)$.
Finally we note that in the gEdS limit, corresponding to $\alpha_2=\alpha_{3AA}=\alpha_{3AB}=\alpha$ 
we obtain 
\begin{eqnarray}
    \nu_2 &=& \frac{2}{3} \Big(\frac{3+14\alpha}{1+6\alpha}\Big),\\
    \nu_3&=& \frac{2}{3} \frac{\left(236 \alpha^2+96 \alpha+9\right)}{(1+6 \alpha)(1+8\alpha)},\\
    \lambda_3 &=&\frac{1}{2}\Big(\frac{1+2 \alpha}{1+8 \alpha}\Big).
\end{eqnarray}

\end{widetext}
\bibliographystyle{apsrev4-2}
\bibliography{references}
\end{document}